\documentclass[prx,aps,floatfix,longbibliography,superscriptaddress,twocolumn,final,notitlepage]{revtex4-2}

\RequirePackage{amsmath}
\RequirePackage{hyperref} 
\RequirePackage{amssymb}
\RequirePackage{graphicx}
\RequirePackage{bbm}
\RequirePackage{xcolor}
\usepackage{natbib}
\usepackage{lipsum}
\usepackage{comment}

\usepackage{xpatch}
\makeatletter
\xpatchcmd{\@ssect@ltx}{\@xsect}{\protected@edef\@currentlabelname{#8}\@xsect}{}{}
\xpatchcmd{\@sect@ltx}{\@xsect}{\protected@edef\@currentlabelname{#8}\@xsect}{}{}
\makeatother

\usepackage{pgf}

\setcitestyle{round}

\usepackage{mathtools}
\usepackage{caption}
\usepackage{subcaption}
\usepackage{amsthm}

\newcommand{\Exp}{\mathbb{E}}
\newcommand{\be}{\begin{equation}}
\newcommand{\ee}{\end{equation}}
\newcommand{\bea}{\begin{eqnarray}}
\newcommand{\eea}{\end{eqnarray}}

\newcommand{\ccup}[1]{\left\{#1\right\}}

\newcommand{\rup}[1]{\left[#1\right]}

\newcommand{\normconst}{\kappa}  

\newcommand{\algoname}{Hy-MMSBM}
\newcommand{\repolink}{\href{http://github.com/nickruggeri/Hy-MMSBM}{github.com/nickruggeri/Hy-MMSBM}}

\DeclareMathOperator{\exppr}{\ell}

\newtheorem{theorem}{Theorem}
\newtheorem{lemma}{Lemma}

\newtheorem*{definition*}{Definition}

\usepackage{amsmath,fixmath}
\usepackage{setspace}

\usepackage[noline, linesnumbered]{algorithm2e}
\RestyleAlgo{ruled}
\SetKwComment{Comment}{> }{}
\SetKwInput{KwInput}{Input}
\SetKwInput{KwYield}{Yield}

\usepackage{float}
\usepackage[nameinlink,capitalize]{cleveref} 

\usepackage{booktabs}
\usepackage{adjustbox}
 
\definecolor{shadecolor}{gray}{0.9}
\usepackage{tabu}

\usepackage[normalem]{ulem} 
\usepackage{soul}

\usepackage{enumitem,amssymb}
\newlist{todolist}{itemize}{2}
\setlist[todolist]{label=$\square$}
\newlist{todolist_done}{itemize}{2}
\usepackage{pifont}
\usepackage{upgreek} 
\let\Omega\upOmega
\let\alpha\upalpha
\let\gamma\upgamma
\let\beta\upbeta
\let\rho\uprho
\let\lambda\uplambda 
\let\kappa\upkappa
\let\theta\uptheta

\DeclareMathOperator{\A}{\mathbf{A}}

\begin{document}

\title[Community Detection in Large Hypergraphs]{Community Detection in Large Hypergraphs}

\author{Nicol\`o Ruggeri}
\email{nicolo.ruggeri@tuebingen.mpg.de}
	\affiliation{Max Planck Institute for Intelligent Systems, Cyber Valley, 72076 Tübingen, Germany}
	\affiliation{Department of Computer Science,  ETH,  8004 Z\"urich, Switzerland}

\author{Martina Contisciani}
	\affiliation{Max Planck Institute for Intelligent Systems, Cyber Valley, 72076 Tübingen, Germany}

\author{Federico Battiston}
	\affiliation{Department of Network and Data Science, Central European University, 1100 Vienna, Austria}

\author{Caterina De Bacco}
	\email{caterina.debacco@tuebingen.mpg.de}
	\affiliation{Max Planck Institute for Intelligent Systems, Cyber Valley, 72076 Tübingen, Germany}

\begin{abstract} 
Hypergraphs, describing networks where interactions take place among any number of units, are a natural tool to model many real-world social and biological systems. In this work we propose a principled framework to model the organization of higher-order data. Our approach recovers community structure with accuracy exceeding that of currently available state-of-the-art algorithms, as tested in synthetic benchmarks with both hard and overlapping ground-truth partitions. Our model is flexible and allows capturing both assortative and disassortative community structures. Moreover, our method scales orders of magnitude faster than competing algorithms, making it suitable for the analysis of very large hypergraphs, containing millions of nodes and interactions among thousands of nodes. Our work constitutes a practical and general tool for hypergraph analysis, broadening our understanding of the organization of real-world higher-order systems.
\end{abstract}

\maketitle

\section{Introduction}
\label{sec:intro}
Over the last decades, most relational data, from biological to social systems, has found a successful representation in terms of networks, where nodes describe the basic units of the system, and links their pairwise interactions~\cite{boccaletti2006complex}. Nevertheless, such a modeling approach cannot properly encode the presence of group interactions, describing associations among three or more system units at a time~\cite{battiston2020networks,torres2021why,battiston2021physics,battiston2022higher}. Such higher-order interactions have been observed in a wide variety of systems, including collaboration networks~\cite{patania2017shape}, cellular networks~\cite{klamt2009hypergraphs}, drug recombination~\cite{zimmer2016prediction}, human~\cite{cencetti2021temporal} and animal~\cite{musciotto2022beyond} face-to-face interactions, and structural and functional mapping of the human brain~\cite{petri2014homological,giusti2016two, santoro2022unveiling}. In addition, the higher-order organization of many interacting systems is associated with the generation of new phenomena and collective behavior across many different dynamical processes, such as diffusion~\cite{carletti2020random}, synchronization~\cite{bick2016chaos,skardal2020higher,millan2020explosive, lucas2020multiorder,gambuzza2021stability,zhang2023higher}, spreading~\cite{iacopini2019simplicial,chowdhary2021simplicial,neuhauser2020multibody} and evolutionary games~\cite{alvarez2021evolutionary,civilini2021evolutionary,civilini2023explosive}. 

Networked systems with higher-order interactions are better described by different mathematical frameworks from networks, such as hypergraphs, where hyperedges encode interactions among an arbitrary number of system units~\cite{berge1973graphs,battiston2020networks}. In the last few years several tools have been developed for higher-order network analysis. These include higher-order centrality scores~\cite{benson2019three, tudisco2021node}, clustering~\cite{benson2018simplicial} and motif analysis~\cite{lotito2022higher, lotito2022exact}, as well as higher-order approaches to network backboning~\cite{musciotto2021detecting, musciotto2022identifying}, link prediction~\cite{contisciani2022principled}, and methods to reconstruct non-dyadic relationships from pairwise interaction records~\cite{young2021hypergraph}.
A variety of approaches have been suggested to detect communities in hypergraphs, including nonparametric methods with hypergraphons \cite{balasubramanian2021nonparametric}, tensor decompositions \cite{ke2019community}, latent space distance models \cite{turnbull2019latent}, latent class models \cite{ng2022model}, flow-based algorithms~\cite{carletti2021random, eriksson2021choosing},
spectral clustering \cite{zhou2006learning,ghoshdastidar2015provable,angelini2015spectral} and spectral embeddings \cite{gong2023generative}. A different line of works focuses on deriving theoretical detectability limits \cite{ghoshdastidar2014consistency,lin2017fundamental,ahn2019community}.

Recently, statistical inference frameworks have been proposed to capture in a principled way the mesoscale organization of hypergraphs~\cite{chodrow2021generative,contisciani2022principled,brusa2022model}. 
Despite their success, current approaches suffer from a number of notable drawbacks. For instance, the method in \cite{brusa2022model} is restricted to utilizing very small hypergraphs and hyperedges, due  to its high computational complexity. Also the approach in \cite{chodrow2021generative} suffers from a high computational complexity in the general case, and needs to make strong assumptions to scale to real-life datasets. Finally, the model in \cite{contisciani2022principled} is constrained to work only with assortative community structures.

In this work we propose a framework to model the organization of higher-order systems. Our method allows detecting communities in hypergraphs with accuracy exceeding that of state-of-the-art approaches, both in the cases of hard and mixed community assignments, as we show on synthetic benchmarks with known ground-truth partitions. Furthermore, its flexibility allows capturing general configurations that could not be previously studied, such as disassortative community interactions. 

Finally, overcoming the computational thresholds of previous methods, our model is extremely efficient, making it suitable to study hypergraphs containing millions of nodes and interactions among thousands of system units not accessible to alternative tools. We illustrate the advantages of our approach through a variety of experiments on synthetic and real data. Our results showcase the wide applicability of the proposed method, contributing to broaden our understanding of the organization of higher-order real-world systems.

\section{Generative model}
A hypergraph consists of a set of nodes $V=\{1, \ldots, N\}$ and a set of hyperedges $E$. Each hyperedge $e$ is a subset of $V$, representing a higher-order interaction between a number $|e|$ of nodes. We denote by $D$ the maximum possible hyperedge size, which can be arbitrarily imposed up to a maximum value of $D=N$, and $\Omega$ the set of all possible hyperedges among nodes in $V$. We represent the hypergraph via an adjacency vector $\A \in \mathbb{N}^{\Omega}$, with entry $A_e$ being the weight of $e \in \Omega$. We assume the weights $A_e$ to be non-negative and discrete. For real-world systems, $\A$ is typically sparse. In fact, the number $|E|$ of non-zero entries is typically linear in $N$, and thus much smaller than the dimension $|\Omega|$. 

We model hypergraphs probabilistically, assuming an underlying arbitrary community structure with $K$ overlapping groups, similarly to a mixed-membership stochastic block model. Each node $i$ can potentially belong to multiple groups, as specified by a $K$-dimensional membership vector $u_i$ with non-negative entries. We collect all the membership assignments in a $N \times K$ matrix $u$. The density of interactions within and between communities is regulated by a symmetric non-negative $K \times K$ affinity matrix $w$. These two main parameters, $u$ and $w$, control the Poisson distributions of the hyperedge weights:
\begin{equation}
\label{eqn: poisson prob hye}
    p(A_e; u, w) = \text{Pois}\left(A_e; \frac{\lambda_e}{\normconst_e} \right) \, ,
\end{equation}
where
\begin{align}
\lambda_e 
    &= \sum_{i<j: i,j \in e}\, u_i^T \, w \, u_j 
    \nonumber \\
    &= \sum_{i<j: i,j \in e}\,\sum_{k,q=1}^K\, u_{ik} \, u_{jq}\, w_{kq}  \, . 
    \label{eqn:lambda}
\end{align}
Here, $\normconst_e = \normconst_{|e|}$ is a normalization factor that solely depends on the hyperedge size $|e|$. 
We develop our theory for a general form of $\normconst_n$. While in principle any choice $\normconst_n > 0$ is possible, in our experiments we utilize the form $\normconst_n = \frac{n (n-1)}{2}\binom{N-2}{n - 2}$, for every hyperedge of size $n$~\cite{ruggeri2022sampling}. Due to the fact that $\normconst_2 = 1$, if the hypergraph contains only pairwise interactions our model is similar to existing mixed-membership block models for dyadic networks \cite{airoldi2008mixed,debacco2017community}. Intuitively, given two nodes $i, j$, the term $\binom{N-2}{n-2}$ normalizes for the number of possible choices of the remaining $n-2$ nodes in the hyperedge. The term $n (n-1) / 2$ averages among the number of possible pairwise interactions among the $n$ nodes in the hyperedge. Note that previous generative models for hypergraphs were limited to detect only assortative community interactions \cite{contisciani2022principled,chodrow2021generative}. By contrast, in our model each entry $w_{kq}$ distinctly specifies the strength of the interactions between each $k, q$ community pair. Hence, for the first time, our method allows encoding more general community structures, without the need to impose a-priori assumptions to ensure computational and theoretical feasibility. In particular, the bilinear form in \cref{eqn:lambda} allows for a tractable and scalable inference, regardless of the structure of $w$. Another relevant feature of the model is that the size of the affinity matrix $w$ does not vary with maximum hyperedge size $D$ nor with the number of hyperedges, making it memory efficient also for hypergraphs with large interactions. We name our model \algoname{}, for Hypergraph Mixed-Membership Stochastic Block Model, and provide an open-source implementation at \repolink{}. We have also incorporated our algorithm inside the open-source library Hypergraphx~\cite{lotito2023hypergraphx}.

\section{Inference}
\label{sec: em inference}
\subsection{Optimization procedure} \label{sec: optimization}
In real-life scenarios, practitioners observe a list of hyperedges, encoded in the vector $\A$, and aim to learn the node memberships $u$ and affinity matrix $w$ that best fit the data.
To this end, we start by considering the likelihood of $\A$ given the parameters $\theta=(u,w)$. Using \cref{eqn: poisson prob hye,eqn:lambda}, this is given by
\begin{equation}
\label{eq: generative model complete}
    p(\A | \theta) = \prod_{e \in \Omega} \text{Pois}\left(A_e; \frac{\lambda_e}{\normconst_e} \right) \, ,
\end{equation}
where the hyperedge weights are assumed to be conditionally independent given $(u,w)$. Its logarithm is given by
\begin{align}
\log p(\A | \theta) 
    &= \sum_{e \in \Omega} -\frac{1}{\kappa_e} \sum_{i< j \in e} u_i^T \, w \, u_j \nonumber \\
    &\hspace{4mm}+ \sum_{e \in E} A_e \log \sum_{i< j \in e} u_i^T \, w \, u_j \,\, , \label{eq:loglik}
\end{align}
where we discarded constant terms not depending on the parameters.
The first summation over $|\Omega|$ terms appears intractable due to the exploding size of the configuration space. However, one important feature of our model is that this high dimensionality can be treated analytically, as the likelihood conveniently simplifies. In fact, the summand $\sum_{e \in \Omega} -\frac{1}{\kappa_e} \sum_{i< j \in e} u_i^T \, w \, u_j$ is simply taking the interaction term $u_i^T w u_j$ as many times as it appears in all the possible hyperedges, each weighted by the factor $1 / \kappa_e$. This reasoning yields the count $C=\sum_{n=2}^D \frac{1}{\normconst_n} \binom{N-2}{n-2}$ and the following simplified log-likelihood:
\begin{align}
\log p(\A | \theta)
    &= - C \sum_{i < j \in V}  u_i^T \, w \, u_j \nonumber \\
    &\hspace{4mm} + \sum_{e \in E} A_e \log \sum_{i< j \in e} u_i^T \, w \, u_j \, , \label{eq:simpifiedLogLik}
\end{align}
obtaining a tractable sum of terms.
To maximize \cref{eq:simpifiedLogLik} with respect to $u$ and $w$, we use a standard variational approach via Jensen's inequality $\log \Exp\rup{x} \geq \Exp\rup{\log x}$, to lower bound the second summand as:
\begin{align}
\hspace{8mm}\sum_{e \in E} A_e \log \sum_{i< j \in e} u_i^T \, w \, u_j &\geq \label{eq: variational ineq}\\
&\hspace{-25mm}\sum_{e \in E} A_e \sum_{i< j \in e} \sum_{k, q =1}^K \rho^{(e)}_{ijkq} \log\left( \frac{u_{ik} \, u_{jq}\,w_{kq}}{\rho^{(e)}_{ijkq} } \right) \, . \nonumber
\end{align}
Here, the variational distribution is specified by the $\rho^{(e)}_{ijkq}$ values, which can be any configuration of strictly positive probabilities such that $\sum_{i< j \in e} \sum_{k, q =1}^K \rho_{ijkq}^{(e)} = 1$.
The equality in \cref{eq: variational ineq} is achieved when
\begin{equation}\label{eqn:rho}
\rho^{(e)}_{ijkq} 
	 = \frac{u_{ik} u_{jq} w_{kq}}{\sum_{i<j \in e} \sum_{k, q=1}^K u_{ik} u_{jq} w_{kq}} 
	 = \frac{u_{ik} u_{jq} w_{kq}}{\lambda_e}  \, .
\end{equation}
Hence, maximizing $\log p(\A | \theta)$ is equivalent to maximizing
\begin{align*}
\mathcal{L}(u,w,\rho) 
    &= - C \sum_{i < j \in V}  u_i^T \, w \, u_j \nonumber \\
    &\hspace{4mm} + \sum_{e \in E} A_e \sum_{i< j \in e} \sum_{k, q =1}^K \rho^{(e)}_{ijkq} \log\left( \frac{u_{ik} \, u_{jq}\,w_{kq}}{\rho^{(e)}_{ijkq} } \right) \quad
\end{align*}
with respect to both $(u,w)$ and $\rho$.
This can be done by alternating between updating $\rho$ and $(u,w)$, as in the Expectation-Maximization (EM) algorithm. \\
The update for $\theta \in \ccup{u,w}$ is obtained by setting the partial derivative $\partial \mathcal{L}(\theta,\rho)/\partial \theta$ to 0, which yields the following expressions:
\begin{align}
u_{ik} &= \frac{
		\sum_{e \in E: i \in e} A_e\rho^{(e)}_{ik}
	}{
	C \sum_{q} w_{kq} \sum_{j \neq i \in V} u_{jq} 
	} \label{eq: u em update}\quad,\\
w_{kq} &=\frac{
		\sum_{e \in E} A_e \rho^{(e)}_{kq}
	}{
		C \sum_{i < j \in V} u_{ik} u_{jq}
	} \label{eq: w em update}\, .
\end{align}
The terms $\rho^{(e)}_{ik}, \rho^{(e)}_{kq}$ are defined as:
\begin{align*}
\rho^{(e)}_{ik} 	&= \sum_{j \in e : j \neq i} \sum_{q } \rho^{(e)}_{ijkq}\, , \\
\rho^{(e)}_{kq}	&= \sum_{i < j \in e} \rho^{(e)}_{ijkq} \, ,
\end{align*}
and obtained after updating $\rho^{(e)}_{ijkq}$ according to \cref{eqn:rho}.
These updates presented in this section are based on maximum likelihood estimation, where we do not set any prior for $(u,w)$. However, we can get Maximum-A-Posteriori estimates (MAP) with similar derivations and complexity by arbitrarily setting priors distributions for the parameters, as we show in Appendix \nameref{sec supp: simplified loglik and MAP}. 
We comment on how to obtain efficient matrix operations that implement the updates in \cref{eq: u em update} and \cref{eq: w em update} in Section \nameref{sec: computational considerations}.

\subsection{Identifiability, interpretation and theoretical implications}
\label{sec: technical considerations}
In the following, we make some observations on relevant aspects regarding the identifiability, interpretation and theoretical implications of the proposed generative model. 
First of all, the log-likelihood in \cref{eq:simpifiedLogLik} is invariant under permutations of the groups and under the rescaling $u \rightarrow c\, u$ and $w \rightarrow w/c^2$, for any constant $c>0$. This observation may raise questions about identifiability of the parameters. However, both permutation and rescaling do not change the composition of the communities nor the relative magnitude of the entries of $w$, thus the mesoscale structure is not impacted by them. Nevertheless, one can easily make the model identifiable by setting a prior probability on $w$ and considering MAP estimates, see Appendix \nameref{sec supp: identifiability} for details. 

Second, for similar invariance reasons, the constant $C$ can be neglected and absorbed after convergence, by either rescaling $u' = \sqrt{C}\, u$ or $w' = C\, w$. While the forms of the rescaling constants $\kappa_e$ play no role during inference, as they only enter the updates through the $C$ term, they do instead impact the generative process when sampling hypergraphs from it~\cite{ruggeri2022sampling}. 
For instance, calculations similar to those in Appendix \nameref{sec supp: avg deg}, allow getting a closed-form expression for the average weighted degree when only considering interactions of size $k$. The resulting formula $\mathbb{E}[d^w_k] = \binom{N-2}{k-2}\frac{k}{\kappa_k \, N} \sum_{i < j \in V} u_i^T \, w \, u_j$ shows that rescaling the constant $\kappa_k$ translates into a rescaling of the average degree. Similar considerations apply to the expected number of hyperedges of a given size, and show that the normalization constants $\kappa_e$ play an important role in determining the expected statistics of the model, and hence of the samples it produces.
Generally, the sampling procedure from the generative model in \cref{eq: generative model complete}, allows determining the degree sequence (i.e. the degree array of the single nodes) as well as the size sequence (i.e. the count of hyperedges for every specified size), which depend on the Poisson parameters and hence on the $\normconst_e$ normalizers. Alternatively the sampling procedure from our generative model can be conditioned to respect such sequences~\cite{ruggeri2022sampling}.

Third, it is possible to obtain the analytical expressions of the expected degree of a node $i$, which evaluates to
\begin{align*}
\Exp\rup{d_i^w} 
    &= \sum_{e \in \Omega: i \in e} \mathbb{E}[A_e] \\
    &= C u_i^T w \,\sum_{j \in V: j \neq i} u_j + C' \sum_{j < m \in V: j, m \neq i} u_j^T w u_m \quad,
\end{align*}
where $C'=\sum_{d=3}^D \frac{\binom{N-3}{d-3}}{\kappa_d}$ is a constant similar to $C$, see Appendix \nameref{sec supp: avg deg}. 
This expression has a relevant interpretation, as it reveals a fundamental difference between simple networks and higher-order systems. Since in dyadic systems $C'=0$, we can think of the rightmost summand as a term contributing only to higher-order interactions, while the leftmost one is a shift of the expected degree coming from binary interactions only. 
One can also observe an analogy with networks of interactions in physical systems. In this context, the leftmost summand can be seen as a mean-field acting on node $i$ in a cavity system where the node is hypothetically removed, while the rightmost term acts as a background field generated by all interactions involving any pair of nodes that does not include node $i$. This background term is peculiar to higher-order systems, as remarked above. Its presence has a relevant effect of building higher-order interactions between nodes in different groups. This can be illustrated with a simple example of a system with assortative $w$ and node $i$ belonging to a different community than all the other nodes. While the leftmost summand yields expected degree zero in dyadic systems, the background field allows $i$ to form on average non-zero edges. Intuitively, this difference is due to the bilinear form in \cref{eqn:lambda}, that allows observing hyperedges that are not completely homogenous, where there could be a minor fraction of nodes that are in different communities than the majority. Notice that such a generation, allowing for mixed hyperedges, is a desirable feature. On the one hand, it is appropriate to model contexts where individuals have multiple preferences and thus are expected to belong to multiple groups. On the other hand, recent work \cite{veldt2021homophily} proves the combinatorial unfeasibility of hypergraphs where all nodes exhibit majority homophily--implying rather uniform hyperedges contained in single communities-- and encourages the development of more flexible generative models.

\subsection{Practical implementation and efficiency}
\label{sec: computational considerations}
From an optimization perspective, the EM algorithm starts by initializing $u$ and $w$ at random and then repeatedly alternating between the \cref{eq: u em update} and \cref{eq: w em update} updates until convergence of $\mathcal{L}(u,w,\rho)$. This does not guarantee to reach the global optimum, but only a local one. In practice, one runs the algorithm several times, each time from a different random initialization, and outputs the parameters corresponding to the realization with highest log-likelihood $\mathcal{L}(u,w,\rho)$. We provide a pseudocode description of the whole inference procedure in \cref{alg: em inference}. For all our experiments, we perform MAP inference on the affinity $w$, setting a factorized exponential prior with rate $1$, and maximum likelihood inference on the assignment $u$. This choice corresponds to the half-Bayesian model presented in Appendices \nameref{sec supp: simplified loglik and MAP} and \nameref{sec supp: identifiability}.
The updates have linear computational cost, obtained by exploiting the sparsity of most real-world datasets with efficient matrix operations, as we show in Appendix \nameref{sec supp: em computations}. Overall, the complexity scales as $O(N\,K + |E|)$, allowing to tackle inference on hypergraphs whose number of nodes and hyperedges was previously prohibitive, see Section \nameref{sec: real data}. 
Another advantage of our inference procedure is that it is stable and reliable for extremely large hyperedges. Due to computational and numerical constraints, previous models were also limited to considering hyperedges with maximal size $D=25$ \cite{chodrow2021generative,contisciani2022principled}. As we illustrate in Section \nameref{sec: real data} with an Amazon and a Gene-Disease dataset, large interactions (respectively $D=9350$ and $D=1074$) should not be neglected as they provide useful information and substantially boost the quality of inference.

\begin{algorithm}
\caption{\algoname{} EM inference}\label{alg: em inference}
\KwInput{Hypergraph $A$, training rounds $r$}
\KwResult{Inferred parameters $(u, w)$} 
\null
BestLoglik = $-\infty$ \\
BestParams = None \\
\Comment{Train model $r$ times and choose \\realization with best likelihood}
\For{$t=1, \ldots, r$}{
    \Comment{Initialize at random}
    $u, w \gets \text{init}(u, w)$ \\
    \Comment{convergence is attained for a max number of EM steps, or below a certain change in parameter values}
    \While{not converged}{
        $u \gets \text{update}(u)$ \hfill \text{\cref{eq: u em update}} 
        \label{algoline: u update}\\
        $w \gets \text{update}(w)$ \hfill \text{\cref{eq: w em update}} 
        \label{algoline: w update}\\
    }
    L = loglik($u, w$) \hfill \cref{eq:simpifiedLogLik} \\
    \If{L \textgreater{} BestLoglik}{
        BestLoglik $\gets$ L \\
        BestParams $\gets (u, w)$ \\
    }
}
\end{algorithm}

\section{Recovery of ground-truth communities}
\label{sec: inference of ground truth communities}
\begin{figure*}
    \centering
    \includegraphics[width=\textwidth]{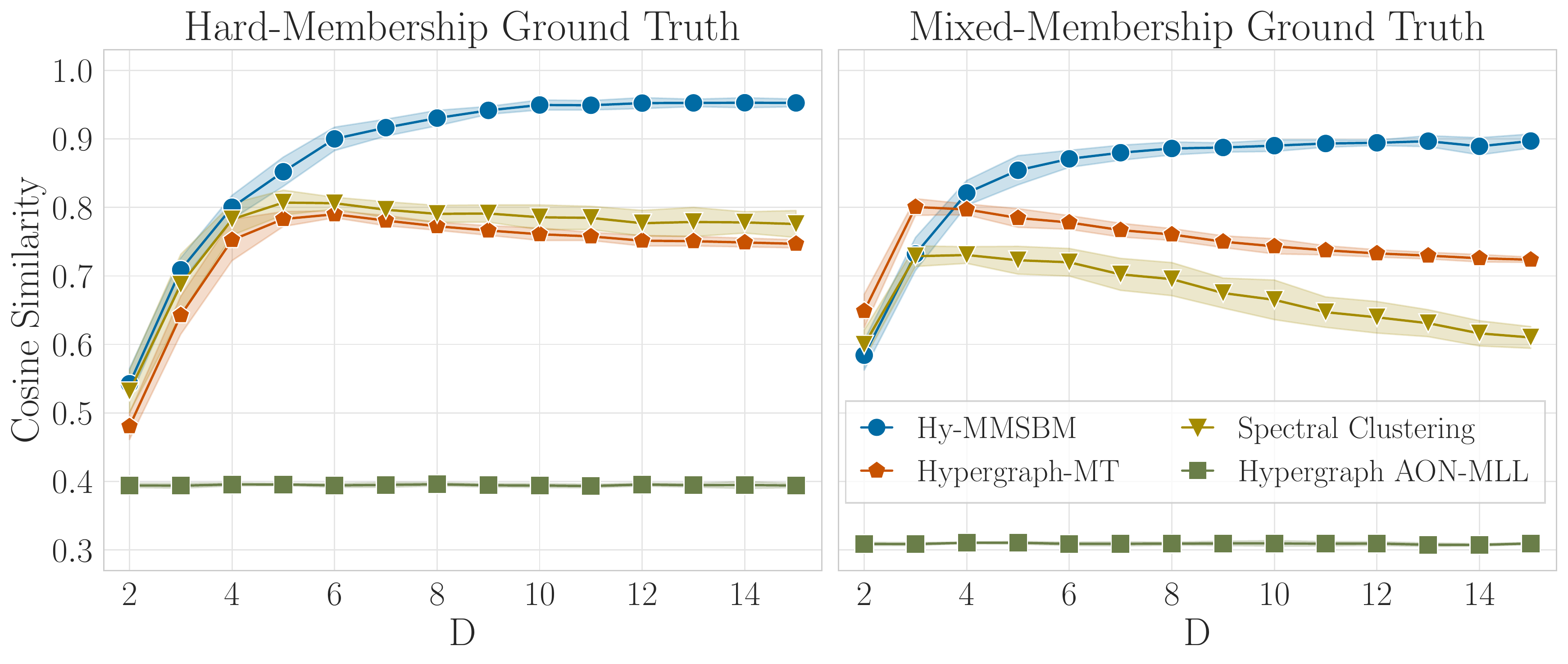}
    \caption{
    \textbf{Recovery of ground truth community assignments.} We measure the cosine similarity between the ground truth and the inferred assignments. We vary the maximum hyperedge size $D$ in synthetic data, and study the cases of hard (left) and mixed (right) ground-truth memberships. 
    When information is scarce, represented by few hyperedges of small maximum size $D$, our method is comparable to the most efficient approaches currently available. However, as larger hyperedges are considered, our method outperforms competing algorithms, both on hard and mixed-membership planted partitions.
    }
    \label{fig:inference of cosine similarity}
\end{figure*}

A standard way to assess the effectiveness of a community detection algorithm is to check if the inferred node memberships match those of a given ground truth. Such ground truth is generally not available for real-world systems~\cite{peel2017ground}, whilst it can be imposed as a planted configuration for synthetic data. 
For this reason, we consider  a recently developed sampling method to produce structured synthetic hypergraphs with flexible structures specified in input~\cite{ruggeri2022sampling}. For further details, see Appendix \nameref{sec supp: data for community detection}. 

In \cref{fig:inference of cosine similarity}, we generate hypergraphs with an underlying diagonal affinity matrix $w$ (assortative structure) and show the recovery performance for the cases of hard (left) and mixed-membership (right) community assignments. The detailed description of the data generation process is provided in Appendix \nameref{sec supp: data for community detection}. We compare our approach with Hypergraph-MT \cite{contisciani2022principled}, an inference algorithm designed to detect overlapping community assignments and assortative interactions; Spectral Clustering \cite{zhou2006learning}, which recovers hard communities via hypergraph cut optimization; and Hypergraph AON-MLL \cite{chodrow2021generative}, which performs a modularity-like optimization based on a Poisson generative model with hard memberships.
For our comparisons, we compute the cosine similarity between the ground truth and the inferred communities, which is appropriate to measure the similarity for both hard and mixed-membership vectors. A value of zero represents no similarity, while a value of one is attained by completely overlapping vectors. 
In both cases, we find that our model successfully recovers the ground-truth communities as more information is made available in terms of hyperedges of increasing sizes. 
This is somehow expected because the generating process of these data reflects the one of our method, and is a sanity check of our maximum likelihood approach. 
Spectral Clustering and Hypergraph-MT attain comparable cosine similarity scores on hard-membership data (left), while their performances differ when detecting mixed memberships (right), with Hypergraph-MT performing better. This is because Spectral Clustering performs an approximate combinatorial search and can only recover hard communities, while Hypergraph-MT allows for overlapping communities via maximum likelihood inference. The low performance of Hypergraph AON-MLL is explained by its generative assumptions. In fact, AON-MLL assigns the same probability to all the hyperedges containing nodes from more than one community. As most of the hyperedges in this synthetic data are made of nodes from more than one community, the recovery of hypergraph modularity on such systems is close to random.
Altogether, such results highlight the effectiveness of the inference procedure, making our model suitable for networked systems with higher-order interactions.
Although relevant, the results in \cref{fig:inference of cosine similarity} are just one possible comparison among algorithms with different generative assumptions. Indeed, such assumptions are expected to yield better or worse results depending on the data, and in general, the no-free-lunch theorem implies that no algorithm will consistently outperform all others on all types of data. As a case for this argument, in Appendix \nameref{sec supp: dcsbm comparison} we present additional results on different synthetic data.

\section{Detectability of community configuration}
\label{sec: assortative vs disassortative experiments}
\begin{figure*}
    \centering
    \includegraphics[width=\textwidth]{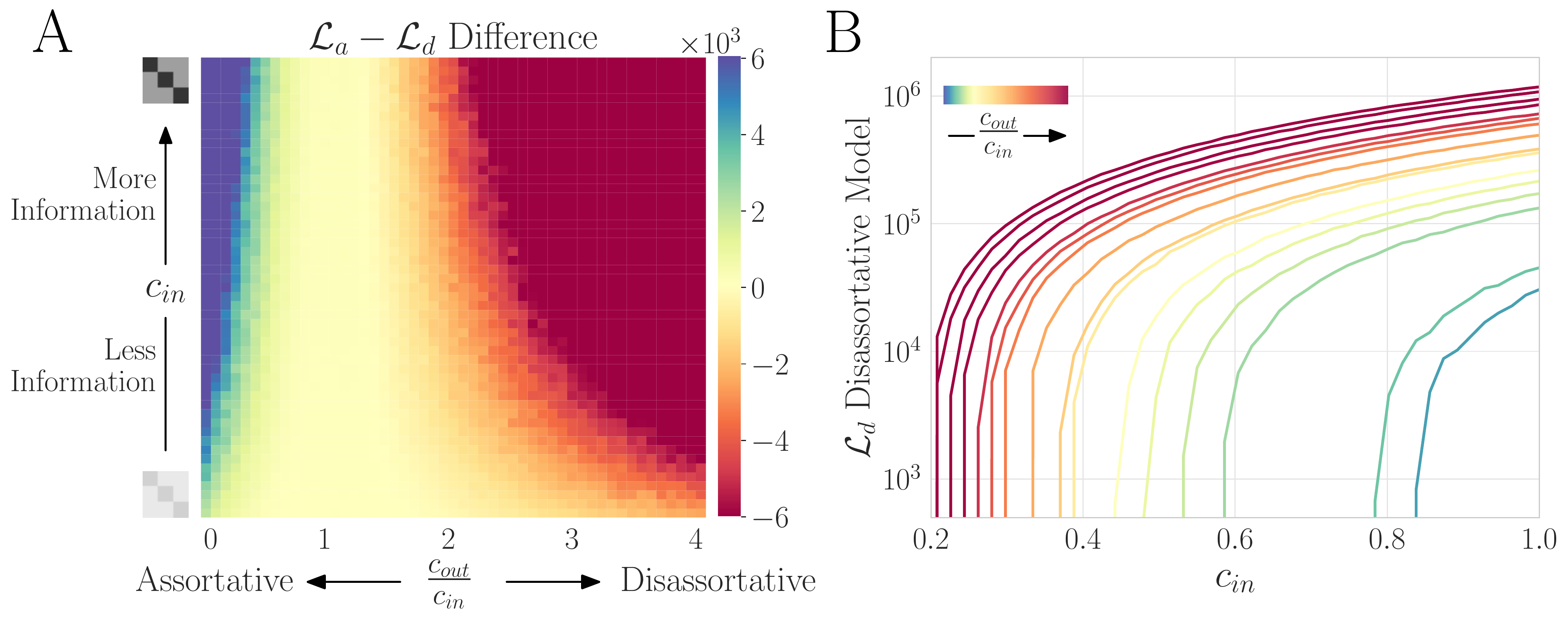}
\caption{\textbf{Detection of assortative and disassortative community interactions}. We generate data where the affinity matrices contain diagonal values $c_{in}$ and out-diagonal $c_{out}$ and measure the ability of our model to detect different assortative and disassortative regimes.
(A) Positive (negative) differences in log-likelihood values indicate that the assortative (disassortative) model attains a better fit. An intermediate regime, highlighted in yellow, also emerges. Here, the detectability is compromised due to not having enough structure ($c_{out} \approx c_{in}$) or enough information (low $c_{in}$).
(B) Log-likelihood of the disassortative model. In this case, the model attains better fit for data with marked disassortative structure (darker red).
}
\label{fig: assortative vs disassortative} 
\end{figure*}

Previous inference algorithms rely on the strong assumption of assortative community interactions, hampering their ability to model more complex mesoscale patterns observed in the real-world. By contrast, our model allows detecting a variety of different regimes, as it assumes a more flexible $w$.

Here, we investigate the detection--and detectability--of different assortative and disassortative community structures in hypergraphs, generalizing previous work on pairwise systems~\cite{decelle2011asymptotic}.
In particular, we generate hypergraphs with hard community assignments, and different community interactions. We take affinity matrices $w$ with diagonal values $c_{in}$ and out-diagonal values $c_{out}$, and vary both $c_{in}$ and the ratio $c_{out}/c_{in}$. By fixing the value of $c_{out}/c_{in}$, we expect higher detectability with increasing $c_{in}$, as this term regulates the expected degree and consequently the information contained in the data. On the contrary, for a fixed value of $c_{in}$, we expect the disassortative model to attain better recovery as the ratio $c_{out}/c_{in}$ increases, due to the stronger inter-community interactions. Details on data generation are provided in Appendix \nameref{sec supp: data for assortative vs disassortative}.

We compare the log-likelihoods obtained by the model when the affinity matrix $w$ is initialized as diagonal or full, which we refer to as \emph{assortative} and \emph{disassortative}, respectively. Notice that the multiplicative updates in \cref{eq: w em update} guarantee that, if $w$ is initialized as diagonal, it will remain as such during training. It is also possible that a full matrix will converge to diagonal during inference. Nonetheless, the strong bias of a diagonal initialization restricts the parameter space of the assortative model, facilitating the convergence to better optima for the detection of assortative structures.

Given the log-likelihood of the assortative ($\mathcal{L}_a$) and disassortative ($\mathcal{L}_d$) models, we measure the difference $\mathcal{L}_a - \mathcal{L}_d$ while varying the values of $c_{in}$ and $c_{out} / c_{in}$. Positive values denote stronger performance of the assortative model, as its likelihood is higher, while negative values favor the disassortative one. We observe that the assortative model attains higher likelihood for low values of $c_{out} / c_{in}$, when within-community interactions are stronger, as shown in \cref{fig: assortative vs disassortative}A. Its performance deteriorates as we increase $c_{out} / c_{in}$ with the disassortative one taking over with higher likelihood values. Furthermore, we can notice an inflexion point at $c_{out} / c_{in} = 1$, where the difference in likelihood between the models is null. While one would expect the disassortative model to perform better in such a scenario, we highlight that this regime is a challenging and noisy one, as the affinity matrix is the uniform matrix of ones. Hence recovery is difficult and not guaranteed, regardless of the model. We finally notice an increase of $\mathcal{L}_a - \mathcal{L}_d$ with $c_{in}$, which regulates the strength of the signal and makes it easier to separate the two regimes.

\begin{figure*}
    \centering
    \includegraphics[width=\textwidth]{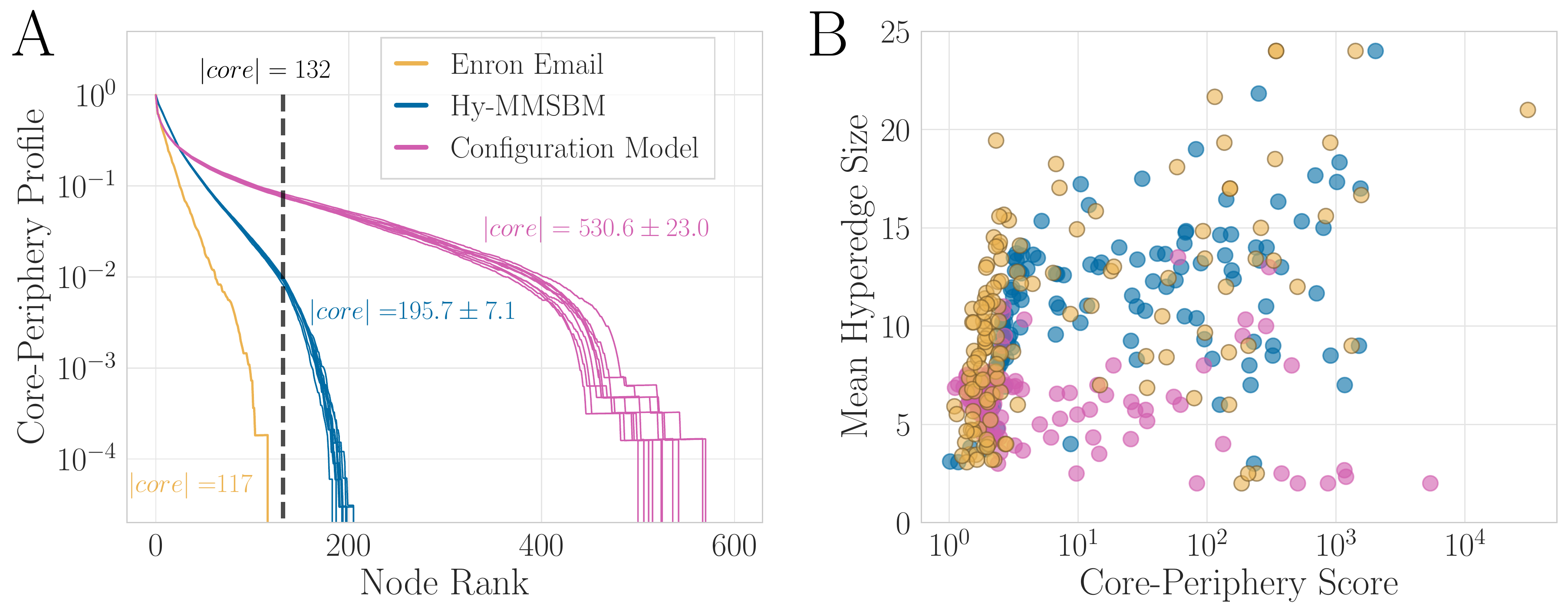}
\caption{\textbf{Recovery of structural core-periphery information}. (A) Core-Periphery profile (\cref{eq:CP-profile}) corresponding to the core-scores computed with HyperNSM on the input Enron email (yellow), ten synthetic samples generated with \algoname\ (blue), and ten synthetic samples generated with a configuration model for hypergraphs (magenta). We plot 600 nodes with the highest core-score in decreasing order, and report the averages and standard deviations of the core dimension for the different datasets. Our method generates samples that closely resemble the property of the input dataset, with an average core dimension close to 132 nodes. (B) Mean size of the hyperedges a node belongs to against its CP score. We observe higher agreement between the data and the inference-based sample generated with \algoname{}. This is also highlighted by the Pearson correlation of the 132 core nodes that is equal to $0.81\pm0.01$ for \algoname{} versus the value of $0.76\pm0.03$ for the samples generated with the configuration model.}
\label{fig: core periphery} 
\end{figure*}

While we expect recovery to improve at more detectable regimes, this may not be observed by only looking at the $\mathcal{L}_a - \mathcal{L}_d$ difference. For this reason, in \cref{fig: assortative vs disassortative}B we complement our analysis by plotting only the log-likelihood $\mathcal{L}_d$ attained via the disassortative initialization. In this case we notice that the performance of the disassortative model increases with both $c_{out} / c_{in}$ and $c_{in}$, as the inter-community interactions get stronger and the expected degree higher. Taken together, our algorithm provides a principled way to extract arbitrary community interactions from higher-order data with varying structural organizations.

\section{Core-periphery structure}
\label{sec: core periphery}
Many real-world systems are characterized by a different mesoscale organization known as core-periphery (CP) structure~\cite{borgatti2000models,cnt016}. Networks characterized by such structure present a group of core of nodes connected among themselves, and often with high degree~\cite{colizza2006detecting,ma2015rich}, and a separate periphery of weakly connected nodes. Recently, methods to study and detect the existence of such patterns in hypergraphs have been proposed~\cite{amburg2021planted, tudisco2022core}. Conceptually, \algoname{} has not been developed with the purpose of core-periphery detection. Nevertheless, we can show its ability in capturing CP structures in hypergraphs through the generation of synthetic data that resemble the core structures of the input dataset.

To measure the recovery of CP structures, we use the method developed by Tudisco et al.~\cite{tudisco2022core}, HyperNSM, that assigns to each node of a hypergraph a core-score quantifying how close the node is to the core, where higher values denote stronger participation. HyperNSM achieved good performance on synthetic and real-world data and its implementation is extremely efficient.

We analyze the Enron email dataset~\cite{klimt2004enron}. Notably, the dataset comes with metadata information identifying a group of core nodes, employees of the organization who send batch emails to the periphery, which in turn only receive emails. This allows us to evaluate the ability of a model to recover a core-periphery structure.
In our study, we utilize the dataset used in Tudisco et al.~\cite{tudisco2022core} with a planted core set that arises directly from the data collection process, as discussed in Amburg et al.~\cite{amburg2021planted} (it is pre-processed by keeping only hyperedges of size $D\leq 25$). The dataset has $N=4423$ nodes and a core composed by $132$ nodes. We apply HyperNSM to quantify the CP structure of the input Enron email dataset, as well as of the samples generated with \algoname{}. To generate the samples, we first run our inference procedure on the Enron email dataset, and then sample hypergraphs distributed according to the obtained $u, w$ parameters. Further details on how to generate the samples are provided in Appendix \nameref{sec supp: core periphery}. For comparison, we also generate samples with a configuration model for hypergraphs~\cite{chodrow2020configuration} and obtain their core-score vectors with HyperNSM as well.

In order to evaluate the quality of the CP assignments in the different samples, we use the CP profile, the metric defined in \cite{tudisco2022core} as: 
\begin{small}
\begin{equation}\label{eq:CP-profile}
    \gamma(S) = \frac{\text{\# hyperedges with \textit{all} nodes in }S}{\text{\# hyperedges with \textit{at least} one node in }S} \, , S \subseteq V \, .
\end{equation}
\end{small}
For any $k \in \{1, \dots, N\}$ we calculate the value $\gamma(S_k(x))$, where $S_k(x)$ is the set of $k$ nodes with smallest core-score in $x$. Given its definition, $\gamma(S)$ is small if $S$ is largely contained in the periphery of the hypergraph and it should increase drastically as $k$ crosses some threshold value $k_0$, which indicates that the nodes in $V \setminus S_{k_0}(x)$ form the core.

In \cref{fig: core periphery}A we show the CP profiles corresponding to the core-scores computed with HyperNSM on the different datasets, i.e. the input Enron email, the samples generated with \algoname, and the samples generated with the configuration model for hypergraphs. We plot 600 nodes with the highest core-score in decreasing order, and for all datasets we notice a sharp drop, which highlights the existence of a CP structure. The main difference is given by the threshold $k_0$ at which this drop happens. This determines the dimension of the core. Remember that the data has a core composed by 132 nodes, and when applying HyperNSM on the input data, we obtain a core dimension equal to 117, validating the good core-detection performance of this algorithm. The samples generated with the configuration model present a core with an average of 530.6 nodes, quite far from what observed in the input dataset. On the other hand, \algoname\ generates samples that better resemble the property of the Enron email dataset, with an average core dimension of 195.7 nodes.

To understand the impact of non-pairwise interactions on higher-order CP structure, we also study the connection between hyperedge size and CP score. In \cref{fig: core periphery}B, we plot the CP score of a given node against the mean size of the hyperedges it belongs to. While we can observe a strong relationship between these two quantities at low CP scores, such regularity disappears in the center of the plot, which contains core nodes and presents a high scattering of hyperedge size values. This unexplained variance is justified by the rich information encoded in the CP score, which jointly depends on different factors related to the topology of the hypergraph.
Yet, the scatter plots obtained on the Enron email dataset and the samples generated with \algoname{}\ have higher similarity than the samples generated with the configuration model. Quantitively, we measure the similarity between the core-scores of the different datasets for the 132 core nodes with the Pearson correlation, a measure $\rho \in [-1, 1]$ of linear correlation between two sets of data. The CP scores of the data have a Pearson correlation equal to $0.81 \pm 0.01$ with the samples generated with \algoname, and of $0.76 \pm 0.03$ with the samples generated with the configuration model.
Similar results are found on the relation between CP score and another structural property, namely the degree of a node, see \cref{fig: cp score vs degree} in Appendix \nameref{sec supp: additional enron}. 

\begin{figure*}
    \centering
    \includegraphics[width=\textwidth]{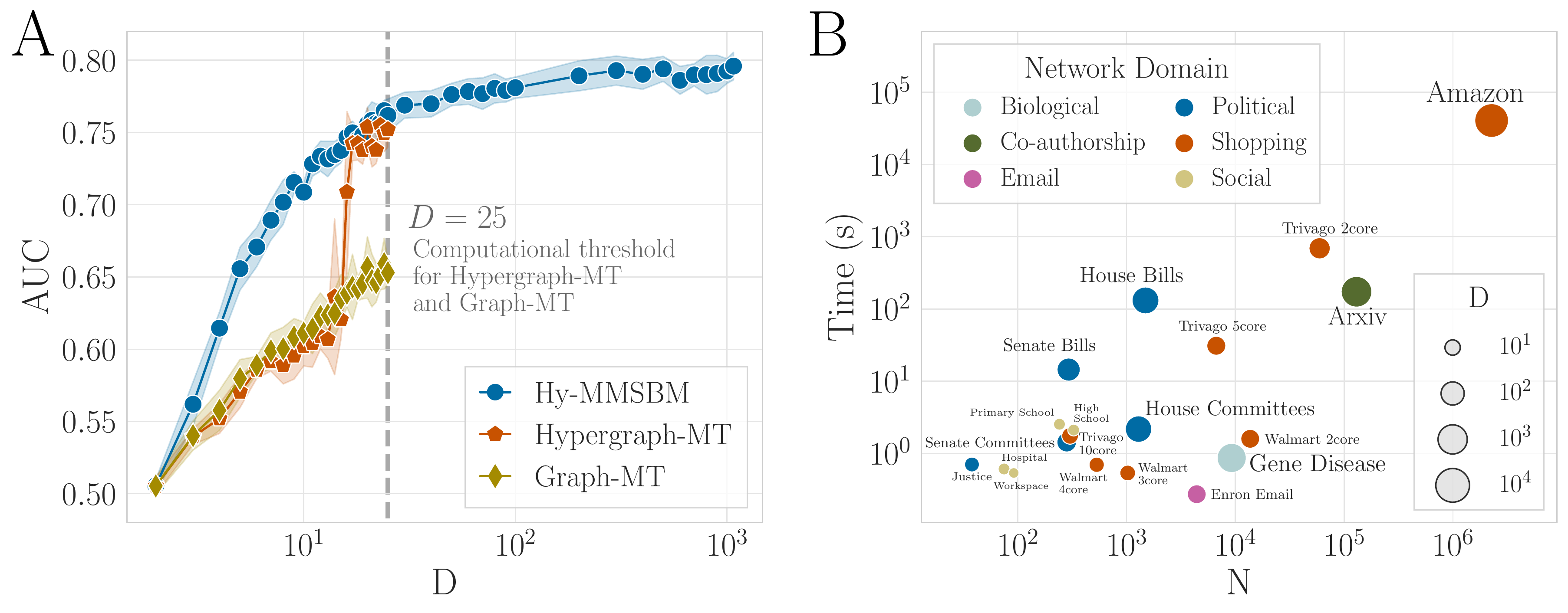}
\caption{
    \textbf{Modeling of real data: hyperedge prediction and running time}. (A) Quality of hyperedge prediction measured by the AUC score on a Gene-Disease dataset, where nodes are genes, and hyperedges contain genes that are associated with a disease. 
    For Hypergraph-MT and Graph-MT the plot shows a computational threshold at the maximum hyperedge size $D=25$. \algoname{} attains the highest scores and is able to model the entire hypergraph, up to $D=1074$.
    (B) Running time of \algoname{} for a variety of real-world datasets. The node represents the data domain. Both $N$ and $D$ are in log-scale. The corresponding AUC scores are reported in \cref{tab: table real datasets}.
    }
\label{fig:test}
\end{figure*}

\section{Modeling of real data}
\label{sec: real data}
In this section, we perform an extensive investigation of higher-order real-world systems. As explained in Section \nameref{sec: em inference} and Appendix \nameref{sec supp: em computations}, the linear-cost EM updates, together with a careful implementation that exploits the sparsity of most datasets, make our method suitable for the analysis of a variety of hypergraphs which were previously inaccessible due to computational constraints. In fact, our method proves to be scalable with respect to both the number of system units and the size of the interactions, improving substantially on competing algorithms currently available in the literature. Moreover, our model is based on a probabilistic formulation, allowing it to perform additional operations and extract information which is not viable via other approaches, such as spectral clustering. First of all, we evaluate the quality of fit of various community detection methods based on their hyperedge prediction capabilities on a Gene Disease dataset, where nodes are genes, and interactions contain genes that are associated with a disease. To this end, we utilize the Area-Under-the-Curve (AUC) measure, a link prediction metric defined as follows: given a randomly selected observed edge, and a randomly selected non-observed one, the AUC $ \in [0,1]$ computes the number of times that the generative model assigns a higher probability to the observed edge. Here, we split the datasets into train and test subsets, where the train sets are used to estimate the parameters, and we evaluate the prediction performance in terms of AUC on the test sets, see Appendix \nameref{sec supp: details experiments real data} for details. Scalability with respect to hyperedge size is a crucial aspect of models for higher-order data. However, due to computational and numerical constraints, previous methods are limited to considering interactions of moderate size only, possibly causing a loss of information and a biased representation of the full system. In contrast, our model is able to efficiently process all the information provided in the dataset, reliably scaling to hyperedges of size of the order of the thousands. In \cref{fig:test}A we compare our method with other probabilistic approaches with hyperedge prediction capabilities. 
When only small interactions are considered, our model outperforms the competitive algorithms. At the computational limit of other approaches $D=25$, Hypergraph-MT and our model attain a similar score, signalling the importance of considering large interactions. Beyond this computational threshold, our method continues to exploit the information provided by interactions among a growing number of units up to the maximum size observed of $D=1074$, which results in an AUC score of $0.79$.   

We then extend our analysis to a variety of datasets from different domains, as described in \cref{fig:test}B. For each dataset we show the inference running time as a function of the number of nodes $N$ and the size of the largest hyperedge $D$. The AUC scores, reported in \cref{tab: table real datasets} and ranging from $0.74$ to $0.98$, show that the model generally yields a good fit and predicts the existence of hyperedges reliably. While these scores are on average aligned with those of other existing algorithms~\cite{contisciani2022principled}, the running time of our model is orders of magnitude lower. This allows studying very large hypergraphs such as the Arxiv, Trivago 2core and Amazon datasets, containing up to millions of nodes and hyperedges. Overcoming the resulting computational challenges, our method allows the efficient modeling of a variety of previously unexplored datasets, which, to the best of our knowledge, could not be tackled by competing higher-order community detection algorithms. 

Taken all together, these results show the effectiveness of our model in tackling datasets of small and large dimensions, both in terms of quantitative performance and computational scalability, and make \algoname{} a valid tool for the study of complex higher-order systems.

\begin{table}
    \centering
    \resizebox{0.96\columnwidth}{!}{%
        \setlength{\tabcolsep}{3pt}
        \renewcommand{\arraystretch}{1.2}
        \begin{tabular}{lrrrrr}
            \toprule
            {} &        $N$ &        $|E|$ &     $D$ &       $K$  &   AUC \\
            \cmidrule{2-6} 
            Justice           &       38 &     2,826 &     9 & 4   & $0.909 \pm 0.008$ \\
            Hospital          &       75 &     1,825 &     5 & 2   & $0.767 \pm 0.013$ \\
            Workspace         &       92 &      788 &     4 & 5   & $0.741 \pm 0.015$ \\
            Primary School    &      242 &    12,704 &     5 & 10  & $0.832 \pm 0.002$ \\
            Senate Committees &      282 &      301 &    31 & 30   & $0.926 \pm 0.023$ \\
            Senate Bills      &      294 &    21,721 &    99 & 13   & $0.921 \pm 0.002$ \\
            Trivago 10core    &      303 &     3,162 &    14 & 11   & $0.960 \pm 0.005$ \\
            High School       &      327 &     7,818 &     5 & 17   & $0.879 \pm 0.007$ \\
            Walmart 4core     &      532 &     2,292 &    10 & 4   & $0.837 \pm 0.013$ \\
            Walmart 3core     &     1,025 &     3,553 &    11 & 4  & $0.825 \pm 0.010$ \\
            House Committees  &     1,290 &      335 &    81 & 25   & $0.939 \pm 0.015$ \\
            House Bills       &     1,494 &    54,933 &   399 & 19   & $0.946 \pm 0.001$ \\
            Enron Email       &     4,423 &     5,734 &    25 & 2   & $0.835 \pm 0.009$ \\
            Trivago 5core     &     6,687 &    33,963 &    26 & 30  & $0.962 \pm 0.001$ \\
            Gene Disease      &     9,262 &     3,128 &  1,074 & 2  & $0.828 \pm 0.010$ \\
            Walmart 2core     &    13,706 &    19,869 &    25 & 2  & $0.788 \pm 0.004$ \\
            Trivago 2core     &    59,536 &   140,698 &    52 & 100 & $0.863 \pm 0.002$ \\
            Arxiv             &   130,024 &   172,173 &  2,097 & 10  & $0.884 \pm 0.001$ \\
            Amazon            &  2,268,231 &  4,242,421 &  9,350 & 29  & $0.978 \pm 0.002$ \\
            \bottomrule
        \end{tabular}
    }
    \caption{\textbf{AUC scores on real datasets}. We report the number of nodes $N$, number of hyperedges $|E|$, maximum hyperedge size $D$, number of communities $K$ and AUC scores attained by our method on 19 large-scale real-world hypergraphs. The results are averages and standard deviations over 10 random test sets, and the value of $K$ is chosen via cross-validation, see Appendix \nameref{sec supp: details experiments real data}.}
    \label{tab: table real datasets}
\end{table}

\section*{Discussion}
In this work we have developed a probabilistic framework to model hypergraphs. Our method allows performing inference on very large hypergraphs, detecting their community structure and reliably predicting the existence of higher-order interactions of arbitrary size. When compared to other available methods on synthetic hypergraphs with known ground truth, for both hard and mixed assignments our model attains the most efficient recovery of the planted partitions. Moreover, compared to previous proposals, \algoname{} relies on less restrictive assumptions on the latent community structure in the data, and is thus able to detect configurations, such as disassortative community interactions, which could not be previously identified.  
Furthermore, our method is extremely fast. Its efficient numerical implementation exploits optimized closed-form updates and dataset sparsity and has linear cost in the number of nodes and hyperedges. 
The resulting formulas are also numerically stable, not resulting in under- or overflows during the computations. Such numerical stability carries over to extremely large interactions, a substantial improvement over the computational threshold of previous methods, allowing to explore higher-order datasets with millions of nodes and interactions among thousands of units, that could not be previously tackled.

There are several directions for future work. 
From a theoretical perspective, our proposed likelihood function is based on a bilinear form for capturing dependencies within the hyperedges, a key ingredient for ensuring both mixed-membership nodes and fast inference. A possible extension would be to consider alternative likelihood definitions where the probability of the hyperedges is determined by multilinear forms, which would in principle allow capturing more complex interactions within the hyperedges.
Similarly, here we have assumed the hyperedges to be independent conditioned on the latent variables.
Relaxing this assumption may ameliorate the expressiveness of the model, allowing to capture topological properties that involve more than two hyperedges, as already observed in the case of networks \cite{safdari2021generative,contisciani2022community,safdari2022reciprocity}.
From an algorithmic perspective, there are different questions that may allow further stabilizing and improving the inference procedure. Among these, the propensity of different initial conditions to be trapped in local optima during EM or MAP inference has not yet been investigated. Devising suitable initialization procedures or parameter priors to favor different membership types, as done in other works~\cite{nakis2023hm}, offers a promising path in this direction. 
Finally, we have considered here a standard scenario where the input data is a list of hyperedges, and these are provided all at once. Other approaches may be needed in case of availability of extra information such as node attributes \cite{newman2016structure,contisciani2020community} or for dynamic data \cite{matias2017statistical}.

Altogether, our work provides an accurate, flexible and scalable tool for the modeling of very large hypergraphs, advancing our ability to tackle and study the organization of real-world higher-order systems.

\section*{Acknowledgements}
\noindent
\textbf{Funding} N.R. acknowledges support from the Max Planck ETH Center for Learning Systems. 
M.C. and C.D.B. were supported by the Cyber Valley Research Fund.
M.C. acknowledges support from the International Max Planck Research School for Intelligent Systems (IMPRS-IS).
F.B. acknowledges support from the Air Force Office of Scientific Research under award number FA8655-22-1-7025.

\noindent
\textbf{Author contributions} 
All authors conceived the project.
N.R. developed the code implementation and performed
the simulations and analysis.
All the authors contributed to the development of models and experiments and to the writing and revision of the paper.
\noindent
\textbf{Competing interests} The authors declare no competing interests.

\noindent
\textbf{Data and Materials availability} All synthetic data needed to evaluate the conclusions of the paper are explained in detail for reproduction. All real data are properly referenced and publicly available.

\bibliographystyle{ScienceAdvances}
\bibliography{bibliography}

\newpage
\null
\newpage
\appendix
\newpage
\onecolumngrid
\section{Maximum-a-Posteriori (MAP) estimation}
\label{sec supp: simplified loglik and MAP}
In this section, we show how to generalize our probabilistic model to include exponential priors on the parameters, and derive the correspondent EM updates for the MAP estimate. 
If we assume factorized exponential priors $p(u_{ik}) \sim \text{Exp}(\exppr_{u_{ik}})$ and $p(w_{kq}) \sim \text{Exp}(\exppr_{w_{kq}})$, where $\exppr_{u_{ik}}, \exppr_{w_{kq}} > 0$ are the rate parameters, then, removing constant terms, we obtain the following log-posterior:
\begin{align*}
\log p(\theta | \A) 
    &= \sum_{e \in \Omega} -\frac{1}{\kappa_e} \sum_{i< j \in e} u_i^T \, w \, u_j  \\
    &\hspace{4mm}+ \sum_{e \in E} A_e \log \sum_{i< j \in e} u_i^T \, w \, u_j \\
    &\hspace{4mm}+ \sum_{i \in V} \sum_{k=1}^K \log p(u_{ik}) + \sum_{k, q=1}^k p(w_{kq}) \, .
\end{align*}
Notice that this expression is given by the log-likelihood in \cref{eq:loglik} plus the prior distribution terms. We can proceed deriving by repeating the calculations in Section \nameref{sec: optimization} to optimize the log-posterior. 
The resulting MAP updates are given by:
\begin{align}
u_{ik} &= \frac{
		\sum_{e \in E: i \in e} A_e\rho^{(e)}_{ik}
	}{
	C \sum_{q} w_{kq} \sum_{j \neq i \in V} u_{jq} + \exppr_{u_{ik}}
	} \label{eq supp: MAP u update}\\
w_{kq} &=\frac{
		\sum_{e \in E} A_e \rho^{(e)}_{kq}
	}{
		C \sum_{i < j \in V} u_{ik} u_{jq} + \exppr_{w_{kq}}
	} \, . \label{eq supp: MAP w update}
\end{align}
Similarly to those for the EM updates in ~\cref{eq: u em update,eq: w em update}, these expressions can be computed efficiently by exploiting the computations presented in Appendix \nameref{sec supp: em computations}, but have the advantage of making the model identifiable, as we explain in Appendix \nameref{sec supp: identifiability}. In our experiments, we utilize default values $\exppr_{u_{ik}} \equiv 0$ and $\exppr_{w_{kq}} \equiv 1$, which correspond to a ``half-bayesian'' model where we fix the relative scale of the parameters via the prior on $w$, and perform frequentist inference on $u$ (as $\exppr_{u_{ik}} \equiv 0$ is not a valid rate for an exponential distribution, but yields the maximum-likelihood EM updates). 

Notice that, in the training procedure, the choice of prior distribution solely impacts the parameter updates. In fact, \cref{eq supp: MAP u update,eq supp: MAP w update} are utilized in \cref{algoline: u update,algoline: w update}  of \cref{alg: em inference}, while leaving the remaining steps unmodified.

\section{Identifiability}
\label{sec supp: identifiability}
In this section, we investigate under which conditions the generative model in \cref{eqn: poisson prob hye,eqn:lambda} is identifiable. As commented in Section \nameref{sec: technical considerations}, the likelihood is invariant to certain simple transformations, such as rescaling or permutations, making the frequentist version of the model non-identifiable. Here, we show how two different modifications, where we impose a prior on only one or both the parameter groups, yield identifiability.

First of all, recall the following definition~\cite{lehmann2006theory}:

\begin{definition*}
Let $p_{\theta}(x)$ be a probabilistic model with parameters $\theta \in \Theta$.  The model is called \emph{identifiable} if, for all $\theta_1, \theta_2 \in \Theta$, the following holds
\begin{equation*}
\theta_1 \neq \theta_2 \implies p_{\theta_1} \neq p_{\theta_2} \, .
\end{equation*}
Notice that two probability distributions $p(x),  q(x)$ on the same space are equal if they differ at most on sets of $x$ values of measure 0.
\end{definition*}

\subsection{Identifiability for the fully Bayesian model}
In the fully Bayesian model we posit both a prior on $u$ and $w$. We prove its identifiability in the following.  
\begin{theorem}\label{th: bayesian idenfiability}
Take the likelihood in \cref{eq:loglik} and posit the following priors for the parameters
\begin{equation*}
\begin{gathered}
p(u_i) \\
p(w_{kq}) \sim \text{Exp}(\exppr) \, ,
\end{gathered}
\end{equation*}
where $p(u_i)$ is any fixed probability distribution with support over $(0, \infty)^K$ and $\exppr$ a valid rate for an exponential distribution. Then the model is identifiable.
\end{theorem}
\begin{proof}
Since the prior on $u$ is fixed, the only parameter in the model is $\exppr$. Take two different values $\exppr^0 \neq \exppr^1$.  Since the joint distribution $p$ has support on the whole space of $u, w$ and $A$ values, the density is non-zero almost everywhere. Thus, we can study the ratio:
\begin{align*}
\frac{p(\A, w, u ; \exppr^0)}{p(\A, w, u ; \exppr^1)}
	&= \frac{
	    p(\A | u, w) p(u) p(w; \exppr^0)
	}{
	p(\A | u, w) p(u) p(w; \exppr^1)
	} \\
	&= \frac{p(w; \exppr^0)}{p(w; \exppr^1)} \\
	&= \prod_{k \le q \in [K]} \frac{\exppr^0 e^{-\exppr^0 w_{kq}}}{\exppr^1 e^{-\exppr^1 w_{kq}}} \\
	&= \left( \frac{\exppr^0}{\exppr^1} \right)^{\frac{K(K-1)}{2}} \hspace{-3mm} \exp\left( (\exppr^1 - \exppr^0) \sum_{k \le q \in [K]} w_{kq}  \right) \, .
\end{align*}
Thus
\begin{equation*}
\frac{p(\A, w, u ; \exppr^0)}{p(\A, w, u ; \exppr^1)} = 1 \iff 
\hspace{-3mm} \sum_{k \le q \in [K]} w_{kq} = \frac{K(K-1) \log\left(\frac{\exppr^1}{\exppr^0}\right)}{\exppr^1 - \exppr^0} \, .
\end{equation*}
The right-hand-side defines a zero-measure subspace of $w$ values. Thus, the two densities are equal only on a subspace of measure zero and the model is identifiable.
\end{proof}

\subsection{Identifiability for the ``half-Bayesian'' model}
While the fully Bayesian formulation of the model is identifiable, it also imposes additional constraints in the form of the priors on both parameters. Here, we show that a ``half-Bayesian'' model, where the affinity values $w$ are thought of as random variables, and the $u$ assignments are kept as frequentist parameters, is still identifiable. Intuitively, the half-Bayesian model resolves the problem of scaling, which is determined via the prior on one set of parameters, without imposing any additional constraints.

\begin{theorem}
Take the likelihood in \cref{eq:loglik} and posit an exponential prior 
\begin{equation*}
    p(w_{kq}) \sim \text{Exp}(\exppr) \, .
\end{equation*}
If $N \ge 3$, the model is identifiable.
\end{theorem}
\begin{proof}
Take two different parameter configurations $(\exppr^0,  u^0) \neq (\exppr^1, u^1)$.  Then there are three cases:
\begin{enumerate}
\item $u^0 = u^1,  \exppr^0 \neq \exppr^1$. The proof proceeds like that of \cref{th: bayesian idenfiability}.

\item  $u^0 \neq u^1,  \exppr^0 = \exppr^1$.  
Since the exponential distribution has non-zero pdf everywhere,  and the likelihood is non-zero for any realization of $A$, we can study the ratio $\frac{p(\A, w; u^0,  \exppr^0)}{p(\A, w ;u^1  \exppr^1)}$.  Call the mean for the hyperedge $e$ given the first parameters $\lambda_e^0 := \sum_{i< j \in e} {u^0_i}^T w u^0_j$ and similarly for $\lambda_e^1$. Then:
\begin{align*}
\frac{p(\A, w ;u^0,  \exppr^0)}{p(\A, w ;u^1  \exppr^1)}
	&= \frac{
        p(\A | w ; u^0) p(w ; \exppr^0)	
	}{
        p(\A | w ; u^1) p(w ; \exppr^1)	
	} \\
	&= \frac{p(\A | w ; u^0)}{p(\A | w ; u^1)} \\
	&= \prod_{e \in \Omega} \frac{Pois(A_e ; \lambda_e^0)}{Pois(A_e ; \lambda_e^1)} \\
	&= \prod_{e \in \Omega} e^{-(\lambda_e^0 - \lambda_e^1)} \left( \frac{\lambda_e^0}{\lambda_e^1} \right)^{A_e} \, .
\end{align*}
By \cref{th: lemma}  there exist some values of $\tilde w,  \tilde e$ such that $\lambda_{\tilde e}^0 \neq \lambda_{\tilde e}^1$ (recall that $\lambda_e^0, \lambda_e^1$ implicitly depend on $w$).  
Assume w.l.o.g. that $\lambda_{\tilde e}^0 > \lambda_{\tilde e}^1$. \\
Choose the following hypergraph realization $\tilde \A$ (implicitly depending on $\tilde w$ through the Poisson parameters $\lambda_e$):
\begin{equation} \label{eq: conditions}
\begin{cases}
    \tilde A_e = 0 
        & \text{if } \lambda_e^0 \le \lambda_e^1  \, .\\
    \null & \null \\ 
      \begin{array}{l@{}}
        \tilde A_e \text{ big enough such that } \\
        e^{-(\lambda_e^0 - \lambda_e^1)} \left( \frac{\lambda_e^0}{\lambda_e^1} \right)^{A_e} > 1
      \end{array}
        & \text{if } \lambda_e^0 > \lambda_e^1 \, . 
\end{cases} 
\end{equation} 
Notice that the second condition is attained for any integer $\tilde A_e > \frac{\lambda_e^0 - \lambda_e^1}{\log(\lambda_e^0 / \lambda_e^1)}$.  With this choice, for every hyperedge $e$ we have $ e^{-(\lambda_e^0 - \lambda_e^1)} \left( \frac{\lambda_e^0}{\lambda_e^1} \right)^{\tilde A_e} \ge 1$. 

Since $\tilde e$ satisfies the second condition of~\cref{eq: conditions}, we obtain the following:
\begin{align*}
\frac{p(\A, w ;u^0,  \exppr^0)}{p(\tilde \A, w ;u^1  \exppr^1)} 
	&= \prod_{e \in \Omega} e^{-(\lambda_e^0 - \lambda_e^1)} \left( \frac{\lambda_e^0}{\lambda_e^1} \right)^{A_e} \\
	&\ge e^{-(\lambda_{\tilde e}^0 - \lambda_{\tilde e}^1)} \left( \frac{\lambda_{\tilde e}^0}{\lambda_{\tilde e}^1} \right)^{\tilde A_{\tilde e}} \\
	&> 1 \, .
\end{align*}

In summary, we found at least one value of $\tilde A, \tilde w$ such that $\frac{p(\tilde \A, \tilde w ;u^0,  \exppr^0)}{p(\tilde \A,  \tilde w ;u^1  \exppr^1)} > 1$. This is not enough to prove that the two probabilities are different: while the set $\{\tilde \A\}$ has non-zero measure (since $\A$ is discrete, every realization has non-zero measure),  the set $\{\tilde w\}$ has zero measure. However observe that $\frac{p(\tilde \A, w ;u^0,  \exppr^0)}{p(\tilde \A, w ;u^1  \exppr^1)} $ is a continuous function in $w$. That means that there exists a neighborhood $U$ of $\tilde w$ where
\begin{equation*}
\frac{p(\tilde \A, w ;u^0,  \exppr^0)}{p(\tilde \A, w ;u^1  \exppr^1)} > 1 \quad \forall w \in U \, .
\end{equation*}
The two probabilities differ on the set of non-zero measure $\{\tilde \A\} \times U$,  and are thus different.

\item  $u^0 \neq u^1,  \exppr^0 \neq \exppr^1$.  
In this case
\begin{equation}
\frac{p(\A, w ;u^0,  \exppr^0)}{p(\A, w ;u^1  \exppr^1)} = 
\frac{p(w; \exppr_0)}{p(w; \exppr_1)} \frac{p(\A | w; u_0)}{p(\A | w; u_1)} \, .
\end{equation}
The proof is the same as in case 2.,  but the value of $\tilde A_e$ in the second condition of~\cref{eq: conditions} is imposed as $\tilde A_e > \frac{\lambda_e^0 - \lambda_e^1 + \log\left(\frac{p(w ; \exppr^1)}{p(w ; \exppr^0)}\right)}{\log(\lambda_e^0 / \lambda_e^1)}$.

\end{enumerate}
\end{proof}

\begin{lemma}
\label{th: lemma}
Suppose $N \ge 3$ and take two values $u^0 \neq u^1$. Then there exist an affinity matrix $\tilde w$ and a hyperedge $\tilde e$ such that $\lambda_{\tilde e}^0 \neq \lambda_{\tilde e}^1$.
\end{lemma}
\begin{proof}
Remember that $u$ is a matrix of shape $N \times K$.  Assume w.l.o.g. that $u_{11}^0 \neq u_{11}^1$ (otherwise we can permute the nodes and community labels such that this is true). \\
By absurd suppose that
\begin{equation}\label{eq: inital absurd}
\lambda_e^0 = \lambda_e^1 \quad \forall w>0,  \forall e \in \Omega \, .
\end{equation}
Notice that relationship in \cref{eq: inital absurd} carries over to all $w \ge 0$,  not only $w>0$. This can be proved as follows. Consider $w \ge 0$ that contains at least one zero value. Then $w + \epsilon > 0$ for any $\epsilon > 0$, and, due to the assumption in \cref{eq: inital absurd}, for any $e$ it is true that $\lambda_e^0(w + \epsilon) = \lambda_e^1(w + \epsilon)$ (where the dependency of $\lambda_e$ on the adjacency matrix has been made explicit).  Then: 
\begin{align*}
\lim_{\epsilon \rightarrow 0^+} \lambda_e^0(w + \epsilon) &= \lim_{\epsilon \rightarrow 0^+} \lambda_e^0(w + \epsilon) \\
\implies \lambda_e^0(w) &= \lambda_e^1(w) \, ,
\end{align*}
by continuity. In summary
\begin{equation}\label{eq: absurd condition}
\lambda_e^0 = \lambda_e^1 \quad \forall w\ge0,  \forall e \in \Omega \, .
\end{equation}
Now consider 
$
w = \begin{pmatrix}
1/u_{11}^0 & \dots & 0  \\
\vdots & \ddots & \vdots \\
0 & \dots & 0
\end{pmatrix}
$.  Relationship \eqref{eq: absurd condition} implies, for any 2-hyperedge $\{1, j\}, \,  j=2, \dots, N$,  that $u_{j1}^0 =  \frac{u_{11}^1}{u_{11}^0} u_{j1}^1$.  Similarly, for any $k \in [K]$, if we define $c_k := \frac{u_{1k}^1}{u_{1k}^0}$, then:
\begin{equation*}
u_{jk}^0 =  c_k u_{jk}^1 \, .
\end{equation*}
or 
\begin{equation}\label{eq: c equivalence}
u_j^0 =  c \odot u_j^1 \, ,
\end{equation}
where $c$ is the vector of $c_k$ values and $\odot$ is the element-wise (or Hadamard) product. 		
Since there are at least three nodes, take some nodes $j, m > 1$. \Cref{eq: absurd condition,eq: c equivalence} imply, for any $w\ge0$ and edge $e=\{j, m\}$ 
\begin{align*}
(u_j^0)^T w (u_m^0) &= (u_j^1) w ^T(u_m^1) \\
\implies (u_j^1 \odot c)^T w (u_m^1 \odot c) &= (u_j^1)^T w (u_m^1) \, .
\end{align*}
By choosing
$
w = \begin{pmatrix}
1 & \dots & 0  \\
\vdots & \ddots & \vdots \\
0 & \dots & 0
\end{pmatrix}
$
we get 
\begin{equation*}
    (u_{j1}^1) c_1^2 = (u_{j1}^1)^2 \implies c_1 = 1 \implies u_{11}^0 = u_{11}^1 \, ,
\end{equation*}
which is a contradiction.
\end{proof}

\section{Computational considerations}
\label{sec supp: em computations}
Naive computations of the EM updates in~\cref{eq: u em update,eq: w em update} incur a quadratic cost $O(N^2)$. Similarly, the computation of the Poisson parameters in \cref{eqn:lambda} could result in expensive operations for large hyperedges. In this section, we show how to simplify the derivations for the Poisson parameters $\lambda_e$ to reach a computational cost of $O(|e|)$, as well as how to make use of sparse data to obtain efficient updates with a linear computational complexity of $O(N)$ and $O(|E|)$.

We utilize a computationally convenient representation of a hypergraph via the incidence matrix $B \in \{0, 1\}^{N \times |E|}$. In every column, indexed by the hyperedges $e$, the incidence matrix contains ones for the nodes in the hyperedge, otherwise zeros, and is typically sparse. For a given hyperedge $e$, we define the quantity $s_e := \sum_{i \in e} u_i$, as well as $s := \sum_{i \in V} u_i$.

\subsection{Computation of $\lambda_e$} 
The following reformulation moves from quadratic to linear computations in $|e|$ for the Poisson parameters $\lambda_e$ defined in \cref{eqn:lambda}: 
\begin{align}
\lambda_e 
	&= \sum_{i< j \in e} u_i^T w u_j  \label{eq supp: lambda naive} \\
	&= \frac{1}{2}(  \sum_{i, j \in e}  u_i^T w u_j - \sum_{i \in e} u_i^T w u_i ) \nonumber \\
	&= \frac{1}{2}( s_e^T w s_e - \sum_{i \in e} u_i^T w u_i ) \, . \label{eq supp: lambda optimized} 
\end{align}
This formulation is efficient for two reasons. First, the cost of the operations move from $O(|e|^2)$ for \cref{eq supp: lambda naive} to $O(|e|)$ for \cref{eq supp: lambda optimized}, which is relevant when performing computations for large hyperedges. Second, we can batch all the calculations to obtain the full vector of $\lambda_e$ values with efficient matrix operations. In fact, the vector $s_e$ is the $e$-th row of $B^T \, u$, resulting in an efficient product between a sparse matrix and a vector, and similarly for the second addend of \cref{eq supp: lambda optimized}.

\subsection{Computation of the $u$ updates}
We recall the following definitions. The Hadamard product between two arbitrarily but equally sized arrays is their elementwise product, which results in a single array with the same shape. For the following derivations, we intend the broadcasting operation between two arrays as their reshaping so that, for every non-matching dimension, a dimension of 1 is added. Every operation we present below is broadcast when dimensions do not match, as is custom for modern scientific computing, and as for all our implementations, which utilize the \textsc{NumPy} and \textsc{SciPy} packages for the \textsc{Python} coding language. We denote the matrix multiplication as \texttt{matmul}. 

The updates in~\cref{eq: u em update} for $u_{ik}$ contain a numerator and a denominator. In the following, we think of $i$ and $k$ as fixed, indexing the matrix of updated $u$ values, and show efficient formulas for the update computations. 

\textbf{Denominator} In the case of the denominator we can obtain the following expression:
\begin{align*}
\sum_{q \in [K]} w_{kq} \sum_{j \neq i \in V} u_{jq}  
    &= \sum_{q \in [K]} w_{kq}(s_q - u_{iq}) \\
    &= \sum_{q \in [K]} w_{kq} s_q - \sum_{q \in [K]} w_{kq}u_{iq} \, ,
\end{align*}
where $s_q$ is the $q$-th element of the $s$ array.
Both addends are batched via matrix multiplications, respectively the matrix-vector product $w \, s$ and matrix-matrix product $(u^T \, w)^T$.

\textbf{Numerator} In the following, call $G \in \mathbb{R}^{N \times |E|}$ the matrix with values $G_{je} := \frac{A_e}{\lambda_e} B_{je}$, which is sparse whenever $B$ is sparse. Then the update formula for $u_{ik}$, can be rewritten as: 
\begin{align*}
\sum_{e \in E: i \in e} A_e\rho^{(e)}_{ik} 
	&= \sum_{e \in E: i \in e} A_e \sum_{j \in e : j \neq i} \sum_{q \in [K]} \frac{ u_{ik} w_{kq} u_{jq} }{\lambda_e} \\
	&= \sum_{e \in E} \frac{A_e}{\lambda_e} B_{ie} \sum_{j \in e : j \neq i} \sum_{q \in [K]} u_{ik} w_{kq} u_{jq} \\
	&= u_{ik} \sum_{e \in E} G_{ie} \sum_{j \in e : j \neq i} \sum_{q \in [K]}  w_{kq} u_{jq} \\
	&= u_{ik} \sum_{e \in E} G_{ie} \sum_{q \in [K]}  w_{kq} (s_{eq} - u_{iq}) \\
	&= u_{ik} 
	\underbrace{\sum_{q \in [K]}  w_{kq} 
		\left(
			\underbrace{\sum_{e \in E} G_{ie} s_{eq}}_{\substack{\small
                    \text{\raisebox{.5pt}{\textcircled{\raisebox{-.9pt} {1}} matmul with}}\\ \text{output } N \times K
                }} - 
			\underbrace{\left( \sum_{e \in E} G_{ie}\right) u_{iq}}_{\substack{\small
                    \text{\raisebox{.5pt}{\textcircled{\raisebox{-.9pt} {2}} broadcast Hadamard}}\\ \text{product with output } N \times K 
                }}
		\right)
	}_{\text{\raisebox{.5pt}{\textcircled{\raisebox{-.9pt} {3}}} matmul with output }N \times K}
	\, .
\end{align*}
Similarly to the case of the Poisson parameters $\lambda_e$, we can interpret this formula in terms of matrix operations, the result being the $(i, k)$ index of the matrix of updated $u$ values. Here, the operations \raisebox{.5pt}{\textcircled{\raisebox{-.9pt}{1}}} and \raisebox{.5pt}{\textcircled{\raisebox{-.9pt}{2}}} yield two $N \times K$ matrices (indexed by $i$ and $q$), whose difference is multiplied with the $w$ matrix in the matrix multiplication \raisebox{.5pt}{\textcircled{\raisebox{-.9pt}{3}}}. The resulting matrix is multiplied element-wise with the $u$ values from the previous iteration. Notice that all the intermediate arrays appearing in the operations consume memory comparable to that of the incidence $B$, and all the matrix multiplications have linear cost in $N$ and $|E|$, resulting in efficient updates. Similar considerations hold for the following section. 

\subsection{Computation of the $w$ updates}
Similar to the $u$ updates above, we look at the numerator and denominator of the $w$ updates in \cref{eq: w em update}. In the following, $k, q$ are considered fixed indices for the matrix of updated values $w$. Once again, notice that all the operations can be performed in time and memory linear in $|E|$ and $N$. 

\textbf{Denominator}
\begin{align*}
\sum_{i < j \in V} u_{ik} u_{jq}  \\
	&\hspace{-10mm} = \frac{1}{2}( \sum_{i, j \in V} u_{ik} u_{jq}  - \sum_{i \in V} u_{ik} u_{iq}) \\
        &\hspace{-10mm} = \frac{1}{2}\left(
        \underbrace{s_k s_q}_{\substack{\text{outer product of } \\ s\text{ with itself}}} - 
        \underbrace{\sum_{i \in V} u_{ik} u_{iq}}_{\substack{\text{matmul of } u\\ \text{ with itself}}}
        \right) \, .
\end{align*}

\textbf{Numerator}
Call $n \in \mathbb{R}^{|E|}$ the vector with elements $n_e := \frac{A_e}{\lambda_e}$ and $F\in \mathbb{R}^N$ the vector $F := B n$. Then:
\begin{align*}
\sum_{e \in E} A_e \rho^{(e)}_{kq} \\
	&\hspace{-10mm} = \sum_{e \in E} A_e \sum_{i< j \in e} \frac{u_{ik} w_{kq} u_{jq}}{\lambda_e} \\
	&\hspace{-10mm} = w_{kq} \sum_{e \in E} n_e \sum_{i< j \in e} u_{ik} u_{jq} \\
	&\hspace{-10mm} = \frac{w_{kq}}{2} \sum_{e \in E} n_e \left( s_{ek} s_{eq} -\sum_{i \in e} u_{ik} u_{iq} \right) \\
	&\hspace{-10mm} = \frac{w_{kq}}{2}  	\left[
    	\sum_{e \in E} n_e s_{ek} s_{eq} - 
    	\sum_{e \in E} \sum_{i \in e} n_e u_{ik} u_{iq}
    	\right] \\
	&\hspace{-10mm} = \frac{w_{kq}}{2}  	\left[
    	\sum_{e \in E} n_e s_{ek} s_{eq} - 
    	\sum_{e \in E} \sum_{i \in V} B_{ie} n_e u_{ik} u_{iq} 
    	\right] \\
	&\hspace{-10mm} = \frac{w_{kq}}{2}  	\left[
    	\sum_{e \in E} n_e s_{ek} s_{eq} - 
    	\sum_{i \in V} u_{ik} u_{iq} \sum_{e \in E}  B_{ie} n_e 
    	\right] \\
	&\hspace{-10mm} = \frac{w_{kq}}{2}  \Bigg[
    	\underbrace{
    		\sum_{e \in E} 
    		\underbrace{n_e s_{ek}}_{\substack{\text{broadcast} \\ \text{Hadamard}}}
    		s_{eq}}_{\substack{\text{matmul with} \\ \text{output } K \times K}} 
    	- 	
    	\underbrace{
    		\sum_{i \in V} u_{ik} 
    		\underbrace{u_{iq} F_i}_{\substack{\text{broadcast} \\ \text{Hadamard}}}
    	}_{\substack{\text{matmul with} \\ \text{output } K \times K}} 
    	\Bigg] \, .
\end{align*}
Notice again that both the appearing Hadamard products are broadcast to match the shapes of the arrays with those of the matrices. Respectively, they are indexed by $e, k$ for the left term $n_e s_{ek}$, and by $i, q$ for the right term $\mu_{iq} F_i$.

\section{Average degree}
\label{sec supp: avg deg}
We now derive the analytical form of the average (weighted) degree of a node. Note that similar calculations were presented in Ruggeri et al.~\cite{ruggeri2022sampling}, we include them here for completeness.
Fix the values of $u, w$. We define the weighted degree $d^w_i$ of a node $i$ as the weighted number of hyperedges it belongs to \cite{battiston2020networks}, i.e. 
\begin{equation*}
d_i^w := \sum_{e \in E: i \in e} A_e = \sum_{e \in \Omega: i \in e} A_e  \, ,
\end{equation*}
where $A_e = 0$ for non-observed hyperedges. We can find its expectation in closed-form:
\begin{align*}
\mathbb{E}[d_i^w]
	&= \sum_{e \in \Omega: i \in e} \mathbb{E}[A_e] 	\\
	&= \sum_{e \in \Omega: i \in e} \frac{\lambda_e}{\kappa_e} \\
	&= \sum_{e \in \Omega: i \in e} \frac{1}{\kappa_e}
		\Bigg(
			\sum_{j \in e: j \neq i} u_i^T \, w \, u_j \\
        & \hspace{2.5cm} + \sum_{j <m \in e: j, m, \neq i} u_j^T \, w \, u_m
		\Bigg) \\
	&= \sum_{e \in \Omega: i \in e} \frac{1}{\kappa_e} \sum_{j \in e: j \neq i} u_i^T \, w \, u_j \\
        &\hspace{4mm} + \sum_{e \in \Omega: i \in e} \frac{1}{\kappa_e} \sum_{j <m \in e: j, m, \neq i} u_j^T \, w \, u_m \\
	&= \sum_{j \in V: j \neq i} 
		\left[
			\sum_{d=2}^D \frac{\binom{N-2}{d-2}}{\kappa_d}
		\right] u_i^T \, w \, u_j \\
	&\hspace{4mm} + \sum_{j < m \in V: j, m \neq i} 
		\left[
			\sum_{d=3}^D \frac{\binom{N-3}{d-3}}{\kappa_d}
		\right] u_j^T \, w \, u_m \\
	&= C u_i^T \, w \sum_{j \in V: j \neq i} u_j \\
        &\hspace{4mm} +  C' \sum_{j < m \in V: j, m \neq i} u_j^T \, w \, u_m \, .
\end{align*}
The step from the fourth to fifth line is justified by the same reasoning we used for reducing the likelihood of the model: we count the number of hyperedges where both node $i$ and $j$ are contained. Notice that we can compute this quantity very efficiently with similar tricks to those from \cref{sec supp: em computations}.

\section{Details for synthetic data generation}
All of our synthetic data are generated by utilizing the method presented in Ruggeri et al.~\cite{ruggeri2022sampling}. Such procedure is based on the same generative model of the inference procedure in \cref{eqn: poisson prob hye}, hence it allows to specify the $u, w$ parameters. Optionally, we condition the samples to respect additional constraints, namely a given size sequence (the number of hyperedges of every dimension) or condition on the sequence of hyperedges observed in the real data.

\subsection{Recovery of community assignments}
\label{sec supp: data for community detection}
In Section \nameref{sec: inference of ground truth communities}, we produce two different datasets, and report the results of different community detection algorithms on such datasets in \cref{fig:inference of cosine similarity}. 
The left-hand dataset, is produced with an assortative matrix $w$ and hard assignments, with $N=500$ and $K=2$ equally sized communities. We condition the samples to the following size sequence, where we specify the number of hyperedges for every size
\texttt{
\{
    2: 500,
    3: 400,
    4: 400,
    5: 400,
    6: 600,
    7: 700,
    8: 800,
    9: 900,
    10: 1000,
    11: 1100,
    12: 1200,
    13: 1300,
    14: 1400,
    15: 1500
\}
}.
This size sequence also implies an average degree varying between $2.0$ (when only considering dyadic interactions) to $248.6$ (when all hyperedges are considered).

The right-hand dataset, is produced with an assortative matrix $w$, $N=500$ and $K=3$ equally sized communities. Differently from the above, the assignments are soft, where the $u_i$ rows are given by $[0.8, 0.2, 0.0]$, $[0., 0.8, 0.2]$ and $[0.2, 0., 0.8]$ for nodes belonging to the three communities, respectively. We condition on the following size sequence 
\texttt{
\{
    2: 700,
    3: 700,
    4: 700,
    5: 700,
    6: 700,
    7: 700,
    8: 800,
    9: 900,
    10: 1000,
    11: 1100,
    12: 1200,
    13: 1300,
    14: 1400,
    15: 1500
\}
}. 
The resulting average degree varies between $2.8$ (for only dyadic interactions) and $257.8$ (when all the hyperedges are considered).

In both cases, we take 10 sampling realizations correspondent to different random seeds. Furthermore, for every seed we extract 10 different samples, which correspond to the same values of degree and size sequence, and are obtained at different time stamps at consequent Markov Chain Monte Carlo steps, as from Ruggeri et al.~\cite{ruggeri2022sampling}. At inference time, only hyperedges up to the specified dimension $D$ are utilized. 

We release part of the synthetic data generated alongside the code that implements our algorithm.

\subsection{Detection of community structure}
\label{sec supp: data for assortative vs disassortative}
Here, we describe the synthetic data generated for the experiments in Section \nameref{sec: assortative vs disassortative experiments}. The data are generated with $K=3$ communities equally split among $N=500$ nodes, maximum hyperedge size $D=20$, and hard assignments $u$. The affinity matrix $w$ has the following form
\begin{equation*}
    w = 
    \begin{pmatrix}
        c_{in}      & c_{out}   & c_{out} \\
        c_{out}     & c_{in}    & c_{out} \\
        c_{out}     & c_{out}   & c_{in} \\
    \end{pmatrix} \, .
\end{equation*}
We vary the value of $c_{in}$ and the ratio $c_{out} / c_{in}$ in a grid with ranges $c_{in} \in [0.1, 1.0]$ and $c_{out} / c_{in} \in [0.0, 4]$.
The expected degrees at the four extremes are: 
\begin{align*}
58.79     &\hspace{1cm}(c_{in}=0.1, \, c_{out} / c_{in}=0.0), \\
531.77    &\hspace{1cm}(c_{in}=0.1, \, c_{out} / c_{in}=4.), \\
587.86    &\hspace{1cm}(c_{in}=1.0, \, c_{out} / c_{in}=0.0), \\
5317.71   &\hspace{1cm}(c_{in}=1.0, \, c_{out} / c_{in}=4.0)  \, . 
\end{align*}

\subsection{Core-periphery experiments}
\label{sec supp: core periphery}
Here, we describe how to generate the data used in Section \nameref{sec: core periphery}. To generate samples as close to the real data as possible, we take an approach similar to experiments for the replication of real data in Ruggeri et al.~\cite{ruggeri2022sampling}, and proceed as follows. For every synthetic sample, we obtain the generative parameters $u, w$ by running the \algoname{} inference procedure on the Enron email dataset. Then, we utilize the inferred values to sample from the generative model. Other than utilizing the inferred parameters, we condition the samples on the hyperedges observed in the real data, which is equivalent to conditioning on the observed size sequence (i.e. the count of observed hyperedges of any given size) and degree sequence (i.e. every node has degree equal to that in the data). For comparison, we also utilize the configuration model of Chodrow~\cite{chodrow2020configuration}. In this case, the sampling is performed by conditioning on the real data's size and degree sequences and mixing the hyperedges through their shuffle-based Markov Chain.

\section{Additional experiments on ground truth recovery}
\label{sec supp: dcsbm comparison}
\renewcommand{\thefigure}{S\arabic{figure}}
\setcounter{figure}{0}
\begin{figure}
    \centering
    \includegraphics[width=0.48\textwidth]{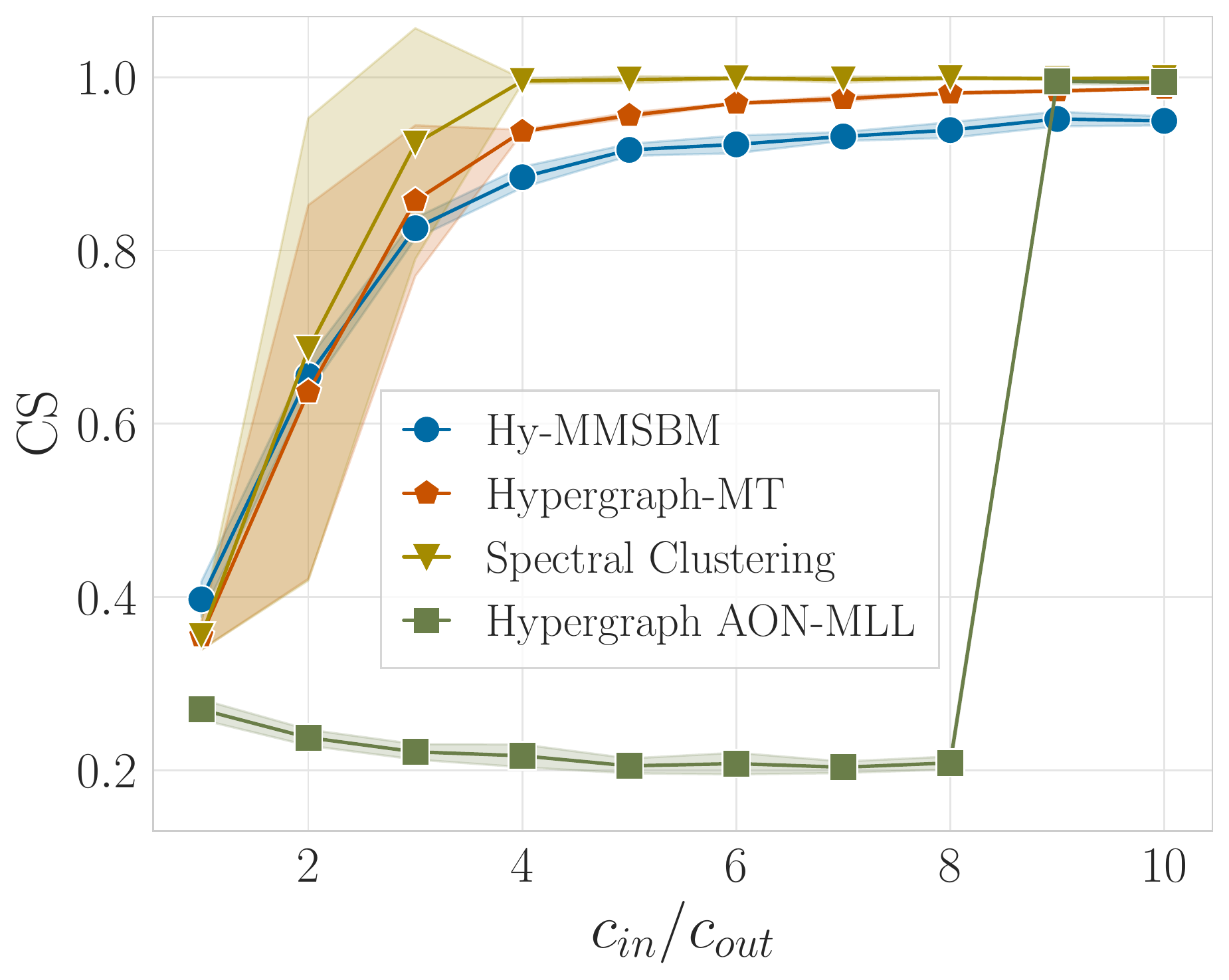}
    \caption{\textbf{Recovery of ground-truth assignments on additional synthetic data}. We show the Cosine Similarity between the ground-truth clusters and the node assignments inferred by different algorithms. On the x-axis, we vary the ratio $c_{in} / c_{out}$ governing the mixing of the hyperedges among different communities. 
    While Hy-MMSBM, Hypergraph-MT and Spectral Clustering all attain good recovery of the planted partitions, their slight differences in performances can be attributed to assumptions more or less aligned with those of the data generating process.
    }
    \label{fig:recovery dcsbm}
\end{figure}
In this section, we present experiments on the recovery of ground truth clusters comparing various algorithms on an additional benchmark dataset. To this end, we produce synthetic data according to the bipartite formalism of Larremore et al.~\cite{larremore2014efficiently} by using the function \texttt{dcsbm\_hypergraph} inside the package \texttt{xgi}~\cite{Landry_XGI}.

We generate hypergraphs with $N = 500$ nodes, $K = 3$ communities, and different values of assortative structure. Specifically, this model takes in input an $\Omega$ matrix that regulates the number of connections within ($c_{in}$) and between ($c_{out}$) communities, and we generate data by varying the strength of the community structure, measured by the ratio $c_{in}/c_{out}$. We fix $c_{in} = 30000$ and vary the ratio $c_{in}/c_{out} \in [1, \dots , 10]$, and for each value we generate 10 independent samples. When $c_{in}/c_{out} = 1$ we have on average $26639.5$ number of hyperedges, an average node degree equal to $536.3$ and the average of the maximum hyperedge size is $33$. On the other extreme, when $c_{in}/c_{out} = 10$, the average number of hyperedges is equal to $21497.1$, the average node degree is $117.2$ and the average maximum hyperedge size is $17.7$. We show the results of comparing various community detection algorithms in \cref{fig:recovery dcsbm}.
\algoname, Spectral Clustering and Hypergraph-MT follow the same pattern, improving their performance as the ratio $c_{in}/c_{out}$ increases. This is expected, as the community structure becomes stronger and easier to detect. Quantitatively, \algoname, Spectral Clustering and Hypergraph-MT perform comparably, and attain good recovery of the planted partitions, with Spectral Clustering performing slightly better than the other methods. Notice that the data are generated with hard community assignments and follow an assortative structure. Therefore, algorithms with stronger inductive biases, such as Spectral Clustering and Hypergraph-MT, tend to perform better. In fact, Spectral Clustering can only detect hard communities, and Hypergraph-MT restricts to assortative configurations while allowing mixed-membership, although, due to its specific likelihood function, with less mixing than \algoname. Despite its flexibility and a generative process different from that of the data, \algoname{} retrieves good results. Hypergraph AON-MLL, instead, poorly performs in all scenarios until the assortative structure becomes dominant ($c_{in}/c_{out} \in \{9, 10\})$, resulting in hyperedges composed by nodes only from one community.

The results in \cref{fig:recovery dcsbm}, together with those presented in Section \nameref{sec: inference of ground truth communities}, allow us to raise an important point about benchmarking, comparing and utilizing different algorithms. In fact, although relevant both in terms of self-consistency and effectiveness of the algorithm, the results in \cref{fig:inference of cosine similarity} can be interpreted under the lens of maximum likelihood consistency, as the generative model of the data corresponds to that of \algoname. Generally speaking, however effective, no algorithm is expected to provide any form of free lunch, performing better than all competitors on all benchmarks. The additional results presented here serve the purpose of showcasing such a trade-off, and of warning practitioners towards the validation of the algorithms employed.

\section{Additional results on the Enron email dataset} \label{sec supp: additional enron}
\begin{figure}
    \centering
    \includegraphics[width=0.48\textwidth]{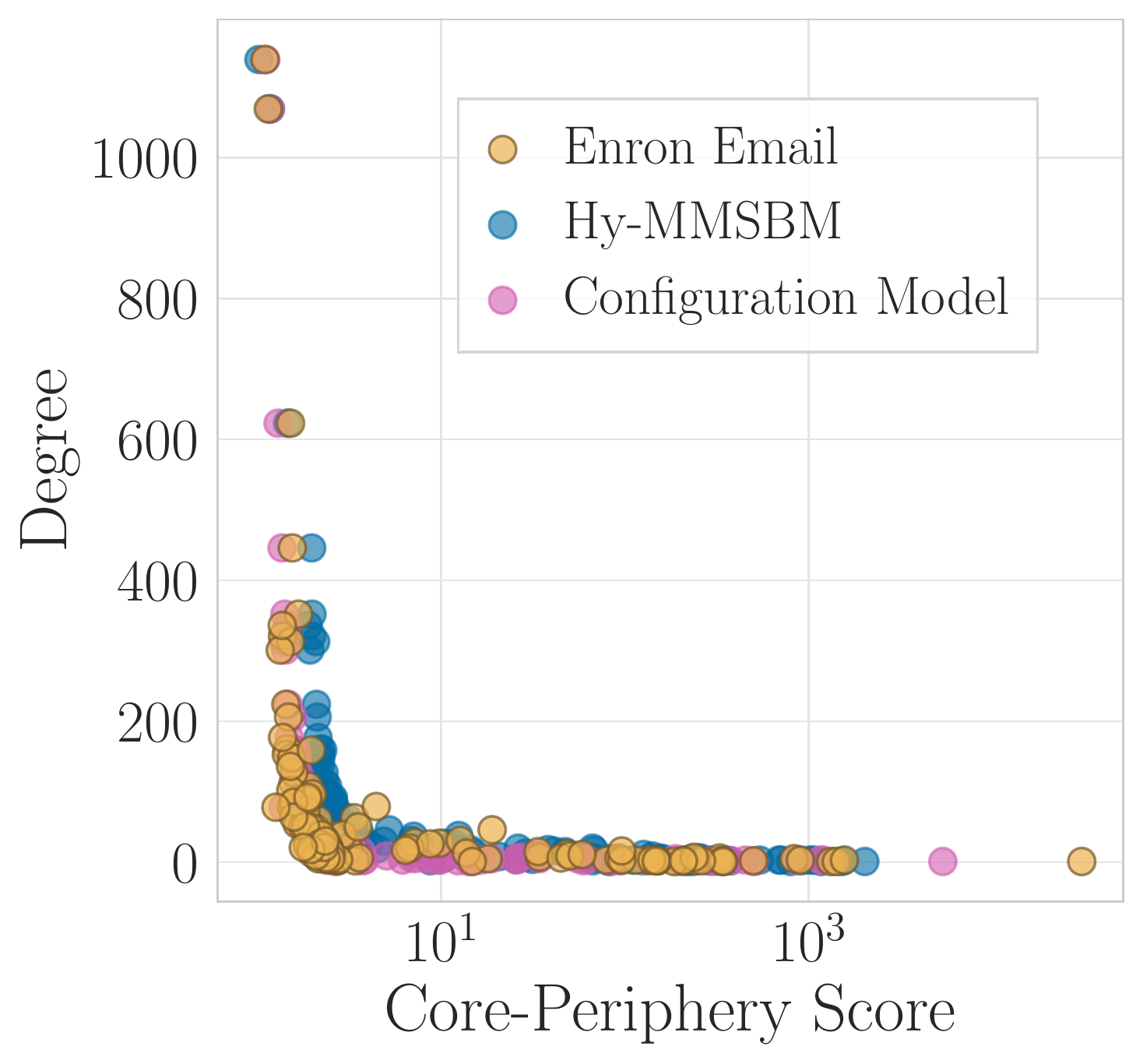}
    \caption{
    \textbf{Recovery of structural core-periphery information}. Additional results on the Enron email dataset~\cite{amburg2021planted}. We plot the degrees of the nodes against their Core-Periphery Scores computed with HyperNSM~\cite{tudisco2022core} on the input Enron email (yellow), one synthetic sample generated with Hy-MMSBM (blue), and one synthetic sample generated with the configuration model for hypergraphs (magenta).  
    }
    \label{fig: cp score vs degree}
\end{figure}

In \cref{fig: cp score vs degree} we show the relationship between the CP score and the degree of a node on the Enron email dataset. Similarly to what observed in Section \nameref{sec: core periphery}, there is a good correspondence between the real data and the samples.

\section{Experiments on real data}
\label{sec supp: details experiments real data}
We report the details for the experiments on real data presented in Section \nameref{sec: real data}.

First, we report on the AUC calculation methodology, and refer to Contisciani et al.~\cite{contisciani2022principled} for further details. To compute the AUC, we split the datasets into a train and test subsets, forming a partition of the hyperedge set. The model is trained on the train set and the reported AUC is computed on the test set. To compute the AUC, we take every hyperedge in the test set and a randomly drawn hyperedge of the same size that is not observed in the dataset, and compare the Poisson probabilities assigned to both by the model. The mean and standard deviations reported in the main text are obtained from 10 models trained with different random seeds. Due to the high computational cost of the AUC calculations, we utilize a ratio of 0.8/0.2 for the train/test partition for all but the Amazon dataset, for which we set a 0.999/0.001 ratio. Notice that the low variance of the AUC scores for the Amazon dataset suggests that the number of hyperedges compared still yields a statistically meaningful calculation. 

Second, we report on the procedure to select the number of communities $K$  for the experiments on real data. The value of $K$ is inferred by computing the AUC scores resulting from a grid of values ranging from $K=2$ to $K=30$, and selecting the one attaining the highest AUC score. For computationally intensive datasets with more than $N=50000$ nodes, namely Trivago 2core, Arxiv and Amazon, we arbitrarily set the value of $K$ to the number of covariates (notice that covariate information is not utilized in any other way). The results of this procedure are the $K$ values presented in \cref{tab: table real datasets}.

\end{document}